\documentclass[]{emulateapj}
\usepackage{amssymb, amsmath}
\usepackage{natbib}
\usepackage{epsfig}
\usepackage{graphicx}

\def\arcmin{\hbox{$^\prime$}}
\def\arcsec{\hbox{$^{\prime\prime}$}}

\shorttitle{NGC~5128 Globular Clusters}
\shortauthors{Sinnott et~al.}

\begin{document}
\title{New \lowercase{$g'r'i'z'$} Photometry of the NGC~5128 Globular Cluster System}
\author{Brendan Sinnott}
\affil{Department of Physics $\&$ Astronomy, McMaster University, Hamilton ON L8S 4M1, Canada}
\email{sinnotbp@physics.mcmaster.ca}
\author{Annie Hou}
\affil{Department of Physics $\&$ Astronomy, McMaster University, Hamilton ON L8S 4M1, Canada}
\email{houa2@physics.mcmaster.ca}
\author{Rachel Anderson}
\affil{Space Telescope Science Institute, Baltimore, MD 21218, USA}
\email{randers@stsci.edu}
\author{William E. Harris}
\affil{Department of Physics $\&$ Astronomy, McMaster University, Hamilton ON L8S 4M1, Canada}
\email{harris@physics.mcmaster.ca}
\author{Kristin A. Woodley}
\affil{Department of Physics $\&$ Astronomy, University of British Columbia, Vancouver BC V6T 1Z1, Canada}
\email{kwoodley@phas.ubc.ca}

\begin{abstract}
We present new photometry for 323 of the globular clusters in NGC~5128 (Centaurus A), measured for the first time in the $g'r'i'z'$ filter system.  The color indices are calibrated directly to standard stars in the $g'r'i'z'$ system and are used to establish the fiducial mean colors for the blue and red (low and high metallicity) globular cluster sequences.  We also use spectroscopically
measured abundances to establish the conversion between the most metallicity-sensitive colors ($(g'-r')_0$, $(g'-i')_0$) and metallicity, [Fe/H].  
\end{abstract}

\section{Introduction}
The metallicity distributions of globular clusters 
(GCs) can provide unique information about the first major episodes of 
star formation that contribute to the formation of the host galaxy
(see \cite{brodie06} and \cite{harris10} for reviews).  
It has now been well 
established that the observed color distribution of old GCs is bimodal in large galaxies of all types,
showing distinct blue and red populations 
\citep{gebhardt99,peng06,brodie06,harris09a}.  The color distribution 
translates into a bimodal metallicity distribution function 
(MDF), 
with the blue clusters corresponding to metal-poor GCs
(with peak [Fe/H]~$\sim-1.5$) and red clusters corresponding
to metal-rich GCs ([Fe/H]~$\sim-0.5$), with a second order dependence on 
galaxy luminosity \citep{brodie06}.  
The accurate conversion from color to metallicity 
is thus an invaluable observational tool for large-scale studies of
GC systems. It can provide a way to construct comprehensive
first-order MDFs based on very large samples (thousands of clusters)
before proceeding with spectroscopic analysis, which requires significantly more telescope time.

Conversions between integrated GC color and metallicity are well known
for metallicity-sensitive broadband color indices, such as $(B-I)$ or $(C-T_1)$, that
have been frequently used in the past \citep[see, e.g.][]{geisler96,harris04b,harris09a}.
Much less work of this type has been done in the US Naval Observatory (USNO) $u'g'r'i'z'$ filter system \citep{smith02} or 
in the similar Sloan Digital Sky Survey (SDSS) $ugriz$ system, but 
the rapidly growing use of these systems 
indicates a need to investigate GC color calibrations in their colors that are
most sensitive to metallicity.  In addition,
the potential for a comprehensive photometric data set over many bands within one galaxy, where
they can be compared very directly, is appealing. 

No single galaxy is an absolutely perfect target
for developing the color/metallicity calibrations. The Milky Way GCs have
the highest quality set of metallicity measurements, but the total GC
population is small and the measurement of their integrated colors requires 
careful large-aperture work; see \cite{peng06} for a published $(g-z)$ calibration
based partly on the Milky Way members.  M31 has a much larger cluster system and
$ugriz$ colors for its GCs have been published \citep{peacock10}, but many of these are affected
by differing and often-uncertain amounts of reddening.  The richest easily accessible
\emph{and} nearly unreddened
collections of GCs are in the Virgo giant ellipticals at $d = 16$ Mpc.  
\citet{jordan09} supply $gz$ photometry of GCs in many Virgo members
and \citet{harris09b} provides $gri$ data for the extremely rich system in M87.
These galaxies, however, lie at larger distances from us so their spectroscopic metallicity measurements 
are far less precise at present than for galaxies in the Local Group.

One of the most attractive individual galaxies for these purposes is NGC~5128, the central giant
in the Centaurus group and the nearest giant elliptical galaxy that can be studied in detail.   
Since its first cluster was identified 
\citep{graham80}, it has been the subject of an extended series
of GC studies \citep[see][for citations and a review]{woodley10b,woodley10a}.
At a distance of 3.8$\pm$0.1 Mpc \citep{gharris09} and moderately low
and uniform reddening ($E_{V-I} = 0.14$) across its halo, NGC~5128 is 
an excellent platform for detailed GC studies, permitting
the investigation of both individual and global properties of the galaxy's 
oldest stellar populations.  The total GC population is estimated to be 
$N_{GC} = 1300 \pm 300$ \citep{gharris10}, of which 607 have now been
individually identified through a combination of
radial velocity measurements \citep{vandenbergh81, hesser84, hesser86, harris92, peng04b, woodley05, rejkuba07, beasley08, woodley10b, woodley10a} 
and resolution into stars
through Hubble Space Telescope (HST) imaging \citep{harris06,mouhcine10}.
In this study, we use an up-to-date catalog of the currently known 
sample \citep{woodley07,woodley10b,woodley10a}\footnote{The two additional GCs not in the Woodley et al.~catalog that were
identified by \citet{mouhcine10} are too faint to appear here.}.  
The purpose of the present paper is to take additional steps towards a calibration of colors
and metallicities in the $g'r'i'z'$ photometric system for this nearby, populous GC system.

The NGC~5128 GC system has now been shown to fall into
the normal pattern of characteristics established from many other
giant galaxies, both elliptical and disk \citep{harris04b,peng04b,peng06}.
The GCs in this galaxy show the standard bimodal color and metallicity distributions 
from both photometry \citep[e.g.][]{harris02ground,peng04b} and spectroscopy
\citep[e.g.][]{peng04b,beasley08,woodley10a}, split roughly 
equally between the metal-poor and metal-rich regimes and with the majority 
being classically old ($> 8$ Gyr).  These studies give every reason to expect
that calibrations of color versus metallicity will be applicable to other galaxies.  Previous large-scale photometric studies 
have been carried out in the normal $UBVRI$ system \citep{peng04a} and
also in the Washington $CMT_{1}$ system \citep{harris92, harris04b, harris04a}.

In this paper we present new measurements for the NGC~5128 clusters in the USNO $g'r'i'z'$ indices, carefully calibrated onto the current standard system.  We then use these measurements, along with previously published spectroscopic data for GCs in both NGC~5128 and the Milky Way, to construct transformations from the $g'r'i'z'$ colors to metallicity, [Fe/H].  In Section~\ref{sec:obsandred}, we discuss our new observations of NGC~5128 and outline our data reduction and photometry procedures, including the creation of our $g'r'i'z'$ catalog of the NGC~5128 GCs. In Section~\ref{sec:diagrams}, we analyze the color-magnitude and color-color diagrams of both the field stars and GCs around NGC~5128, and in Section~\ref{sec:calib} we discuss the results of our color-metallicity calibration.  In Section~\ref{sec:filtersystems}, we discuss the use of the $g'r'i'z'$ filter system for future GC studies.  We conclude with our results in Section~\ref{sec:conclusions}.

\section{Observations and Reductions}
\label{sec:obsandred}
Imaging of the NGC 5128 field was taken over five consecutive
nights in May 2008 with the Yale/SMARTS 1.0m telescope at the {\it Cerro Tololo Inter-American Observatory}, Chile. 
Photometry was performed with the 20$\arcmin$~$\times$~20$\arcmin$
Y4KCam imager in the SDSS $griz$~$\approx~g'r'i'z'$ filters\footnote{Transformations from 
\citet{tucker06} show the mean differences $\Delta m=m(\text{USNO})-m(\text{SDSS})$ 
for GCs observed in this study are $\Delta (g,r,i,z) = (-0.012,-0.005,-0.006,0.004)$.}. 
The scale of the camera is $0.289\arcsec$px$^{-1}$, and at
the 3.8 Mpc distance of NGC~5128, $1'$ corresponds to a linear scale of 1.1 kpc.
During the observing run, a malfunctioning amplifier rendered the NE quadrant of the detector unusable. 
We compensated by using three overlapping pointings, with NGC~5128 positioned on the central, western, and southern 
regions of the detector. Observations covered a 20$\arcmin$~$\times$~20$\arcmin$ region centered on NGC~5128 and 
two 10$\arcmin$~$\times$~10$\arcmin$ regions east and north of the galaxy. Creating an overlapping grid on the sky, 
we were able to recover most of our originally planned spatial coverage of the NGC~5128 halo. 
The total integration time for each pointing was 0.8-1.4 hours in each filter. 
The seeing ranged from FWHM 1.1$\arcsec$ to 1.8$\arcsec$ over the course of the run\footnote{We note that for 
the Yale/SMARTS 1.0m telescope, the observed FWHM is typically 
$\sim$~0.5$\arcsec$ higher inside the dome than the outside seeing recorded by the dimm monitor.}.

For standardization purposes, 14 $u'g'r'i'z'$ standard stars from \citet{smith02} were also observed every night over the 
course of the run, typically resulting in 20 independent integrations each night, at intervals of 2--4 hours and
over a wide range of colors and airmasses.

\subsection{Data Reduction}
Master flat fields for each filter were constructed from a combination
of twilight flat exposures and dome flats.  In addition, on-sky ``blank fields'' 
(high-latitude star fields devoid of bright galaxies and with minimal populations of field stars) 
were observed each night and combined to construct a final
illumination correction for the camera.  The flat-fielding plus illumination correction allowed us to correct for the
sensitivity across the detector to within 1$\%$.

All frames were trimmed, overscan-subtracted, bias-corrected, and flat-fielded with the Massey 
Y4KCam scripts\footnote{http://www.lowell.edu/users/massey/obins/y4kcamred.html} 
written for IRAF\footnote{IRAF is distributed by the National Optical Astronomy Observatories, 
which are operated by the Association of the Universities 
for Research in Astronomy, Inc., under cooperative agreement with the National Science Foundation.}. 
Illumination corrections were applied to the $g'$ and $r'$ images. 
In the $i'$ and $z'$ filters, fringing patterns were also removed by constructing master fringe frames from smoothed medians of the blank-field exposures.  These were subtracted from our science frames once normalized to their exposure time.

Large-aperture photometry was performed on the standard stars
with a 13 pixel (3.78$\arcsec$) radius, determined via a 
curve-of-growth analysis \citep[][and the \emph{digiphot.appphot} IRAF package]{stetson90}.
Nightly photometric calibrations were derived from our observations of the standard stars
and the catalog magnitudes in \citet{smith02}, along with a
linear model for the transformations
\begin{eqnarray}
g'_{\text{inst}} =& g'_{\text{std}} + g'_{0} + k_{g'} X_{g'} + a_{g'}(g' - r')_{\text{std}}\label{g}\\
r'_{\text{inst}} =& r'_{\text{std}} + r'_{0} + k_{r'} X_{r'} + a_{r'}(r' - i')_{\text{std}}\label{r}\\
i'_{\text{inst}} =& i'_{\text{std}} + i'_{0} + k_{i'} X_{i'} + a_{i'}(i' - z')_{\text{std}}\label{i}\\
z'_{\text{inst}} =& z'_{\text{std}} + z'_{0} + k_{z'} X_{z'} + a_{z'}(i' - z')_{\text{std}}\label{z}
\end{eqnarray}
where $X$ represents the airmass and $g'_{inst}$, etc.~are 
the large-aperture magnitudes on the internal instrumental scale.
The color coefficients, $a_x$, were adopted to be constant over
the five consecutive nights of the run; our least-squares solutions gave mean values
\begin{eqnarray*}
a_{g'} =&0.003\pm0.008\\
a_{r'} =&-0.016\pm0.009\\
a_{i'} =&-0.021\pm0.011\\
a_{z'} =& 0.015\pm0.026
\end{eqnarray*}	 
These color terms are all nearly zero and verify that the Y4KCam SDSS filters 
provide a very close match to the standard Sloan system.  The airmass coefficients, $X_x$, were solved
for each night.  Since the airmasses were primarily in the range $X \sim 1.0 - 1.4$, our results are rather
insensitive to the precise $k_x$ values.

We then carried out the photometry for all objects in our NGC~5128 
program fields using \emph{daophot} and \emph{phot} small-aperture photometry 
within the \emph{digiphot.daophot} package.  
Bright isolated stars
were used to construct a mean point spread function (PSF) for
each field and each filter.
With observations taken in $>$1.1$\arcsec$ seeing conditions, most of the GCs in NGC~5128 appear 
very nearly star-like \citep[a normal half-light diameter of 5 pc is equivalent to $0.27''$; see also][]{harris02ground},
enabling us to determine their relative
magnitudes and colors from PSF-fitting photometry within \emph{digiphot.daophot.allstar}. 
PSF magnitudes were converted to the same large-aperture magnitude scale by a mean 
offset $\langle \Delta m \rangle = m_{\text{large ap}}-m_{\text{PSF}}$ determined
from bright, isolated target objects. 

The World Coordinate System (WCS) solutions for the astrometry were derived in two steps.
An initial guess to the solution was found with 
the public-domain astrometric calibration program \emph{astrometry.net}
described in \citet{lang10}.  The first-order
WCS solution was fed into WCSTOOLS \citep{mink02}, where a 
precise solution was found by matching to the USNO UCAC2 catalog.  We find that the resulting WCS solutions have
mean uncertainties less than $0.25\arcsec$ for all measured objects.

\subsection{The NGC~5128 Globular Cluster Catalog}
\label{sec:catalog_GCs}
A photometric catalog of NGC~5128 sources was created for each night of observing, as well as each pointing. We first removed all 
stars landing in the dead NE quadrant of the CCD from our photometry lists. The objects included in the catalog
are those which were detected and measured in \emph{both} the $r'$ and $i'$ filters
(though not necessarily the $g'$ or $z'$ filters, which had slightly shallower detection limits 
for objects of intermediate colors like those of GCs). 
We then determined the $g'r'i'z'$ magnitudes on the USNO system for all sources for each night and 
field pointing  with the 
$\langle \Delta m \rangle$ offset and the inverse of the transformation equations given above. 
Since the inverse of Equations~\ref{g}-\ref{z} depend strictly on $z'$, we used an appropriate
average color of $(i'-z')_{\text{avg}}=0.277$ (see Figure~\ref{fig:ccdsall} below) to solve for the $g'$, $r'$, and $i'$ apparent 
magnitudes in cases where the $z'$ magnitude was not well defined.

The catalogs for each separate night were combined by matching objects within a 
globally selected radius such that all objects have a unique match.  
The astrometry and photometry was then averaged, weighting the 
photometry with the inverse of the square of the measurement uncertainty. 
In cases where the photometry of the same object on different nights differed by more than 0.15 mag, the magnitude with the smaller uncertainty
was selected. 

\begin{figure}[hbt!]
\centering
\includegraphics[width = 8cm, height = 8cm]{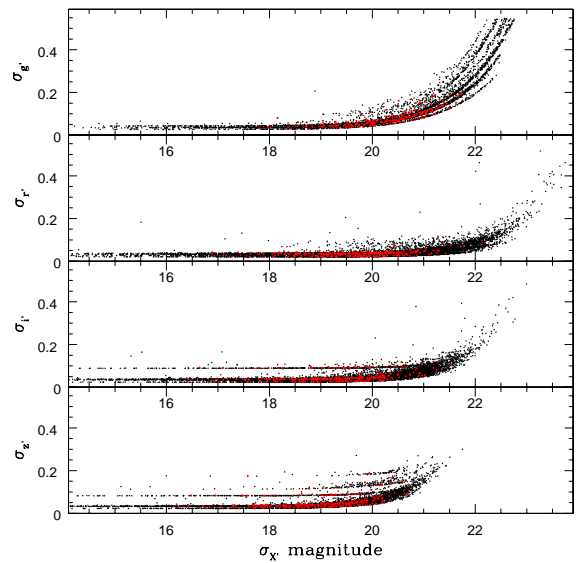}
\caption[]{From top to bottom, photometric uncertainties as a function of magnitude for the $g'$, $r'$, $i'$, and $z'$ filters. 7,026 sources measured in the NGC~5128 field are shown, with matched GCs over-plotted in red (visible in the online edition). Sources with systematically larger 
uncertainties (the thin sequences sitting above the main populations) correspond to objects in regions not overlapped by the pointings, and thus 
measured once.}
\label{fig:magerrs}
\end{figure}

Our final catalog of NGC~5128 sources was created from all stars in each pointing. 
The three major pointings partially overlapped, and for the areas of overlap we again apply a weighted average. 
The final catalog contains 7026 sources for which at least $r'$ and $i'$ photometry is measured. 
The uncertainties for the NGC~5128 photometry are shown in Figure~\ref{fig:magerrs}.
The majority of these sources are a mixture of foreground Milky Way stars and faint
background galaxies, so the first step in the following analysis is to select out the GCs that genuinely
belong to NGC 5128.

\begin{deluxetable*}{cccccccccccc}[hbt!]
\centering
\tabletypesize{\scriptsize} 
\tablecaption{ $g'r'i'z'$ Photometry Catalog Of NGC~5128 GCs
\label{tbl:GCgriz}}
\tablehead{ 
\colhead{ID} & \colhead{$\alpha$} & \colhead{$\delta$} & \colhead{V} & \colhead{$g'$} &  \colhead{$\sigma_{g'}$} & \colhead{$r'$} &  \colhead{$\sigma_{r'}$} & \colhead{$i'$} &  \colhead{$\sigma_{i'}$} & \colhead{$z'$} &  \colhead{$\sigma_{z'}$}\\ \colhead{} & \colhead{(J2000)} & \colhead{(J2000)} &
} 
\startdata
GC001 & 00:53:40.08 & -42:56:51.995 & 18.75 & 19.255 & 0.033  & 18.477 & 0.025  & 18.086 & 0.034  & 17.858 & 0.031 \\
GC039 & 00:53:38.58 & -43:06:26.660 & 19.45 & 20.046 & 0.059  & 19.129 & 0.034  & 18.683 & 0.038  & 18.417 & 0.037 \\
GC043 & 00:53:38.70 & -42:53:35.130 & 99.00 & 99.000 & 99.00  & 21.718 & 0.068  & 21.188 & 0.091  & 99.000 & 99.00 \\
GC045 & 00:53:38.74 & -42:59:48.380 & 20.23 & 20.439 & 0.086  & 19.891 & 0.047  & 19.442 & 0.048  & 19.137 & 0.046 \\
GC046 & 00:53:38.75 & -43:01:45.605 & 19.22 & 19.708 & 0.054  & 19.017 & 0.038  & 18.722 & 0.042  & 18.552 & 0.039 \\
... & ... & ... & ... & ... & ... & ... & ... & ... & ... & ... & ...\\
  \enddata
\tablecomments{Table~\ref{tbl:GCgriz} in its entirety can be found in the Appendix.}
\end{deluxetable*}
We used the GC database from \citet{woodley07,woodley10b,woodley10a} to match 605 confirmed GCs in NGC~5128 to our catalog of $g'r'i'z'$
measurements. 
To obtain the most complete set of correlations, we tried matches out to a search radius of $5\arcsec$,
although we found that $\sim95\%$ of the matches were unique to within
$0.8\arcsec$ with an average angular separation of $0.25\arcsec$, consistent with the astrometric accuracy
described above. Any remaining duplicates that were clearly false matches
were eliminated through their very different colors or magnitudes.
Our final list contains $g'r'i'z'$ photometry for 
323 confirmed GCs of NCG~5128, representing $\sim80\%$ of the known GCs within our field of view. 
The median uncertainty in the GC photometry for the $g'$, $r'$, $i'$, and $z'$ filters is 0.069, 0.035, 0.040, and 0.040, respectively. 
A truncated version of the catalog 
is given in Table~\ref{tbl:GCgriz}, with the entire listing available in the online edition.  The successive columns give the ID number from 
\citet{woodley07,woodley10b,woodley10a}, J2000 coordinates from these studies,
the $V$ magnitude from \citet{peng04b}, and the $g'r'i'z'$ magnitudes with their internal uncertainties.
Any ``99.00'' values indicate no data.


\section{Color-Magnitude and Two-Color Diagrams}
\label{sec:diagrams}

\begin{figure}[hbt!]
\centering
\includegraphics[width = 8.1cm, height = 8.1cm]{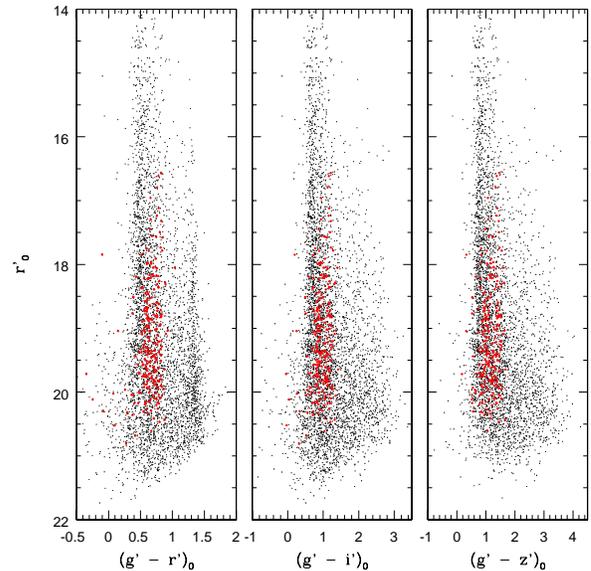}
\caption[]{Color-magnitude diagrams for all measured NGC~5128 sources,
with matched GCs over-plotted in red points (visible in the online edition).}
\label{fig:cmdsall}
\end{figure}

\begin{figure}[hbt!]
\centering
\includegraphics[width = 8.1cm, height = 8.1cm]{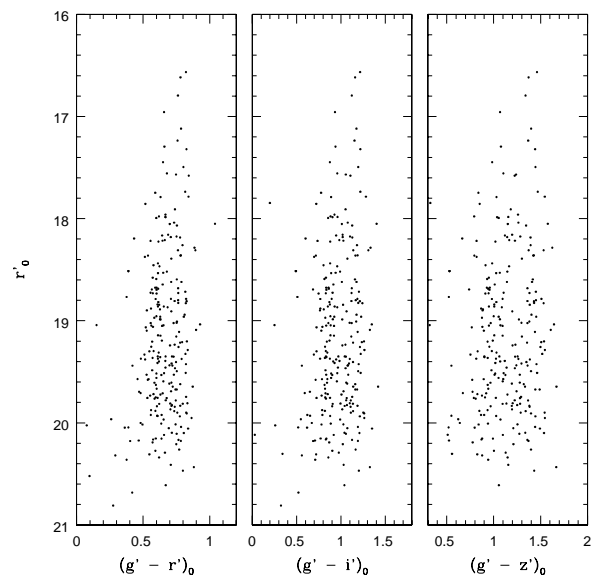}
\caption[]{Color-magnitude diagrams of the 323 matched GCs in NGC~5128. }
\label{fig:cmdsgcs_vsr}
\end{figure}

Figure~\ref{fig:cmdsall} shows the color-magnitude diagrams of all measured objects
in our NGC 5128 field
in the form $r'_0$ versus $(g'-r')_0$, $(g'-i')_0$, and $(g'-z')_0$. 
The GCs are over-plotted in red. The photometry is corrected for foreground reddening 
with $A_r' = 0.879\cdot A_V = 0.33$, $E_{g'-r'} = 0.10$, $E_{r'-i'}=0.07$, $E_{i'-z'}=0.04$
\citep{fukugita96}.  As noted above, about 95\% of the objects in this diagram are 
field contamination, with most concentrated along $(g'-i')_0 \sim 1$.
In Figure~\ref{fig:cmdsgcs_vsr} we show the color-magnitude diagrams for the GCs only. 
The characteristic bimodal color distribution is visible in all three plots, although is
most obvious in the metallicity-sensitive $(g'-i')_0$ and $(g'-z')_{0}$ indices.

\begin{figure}[hbt!]
\centering
\includegraphics[width = 10cm]{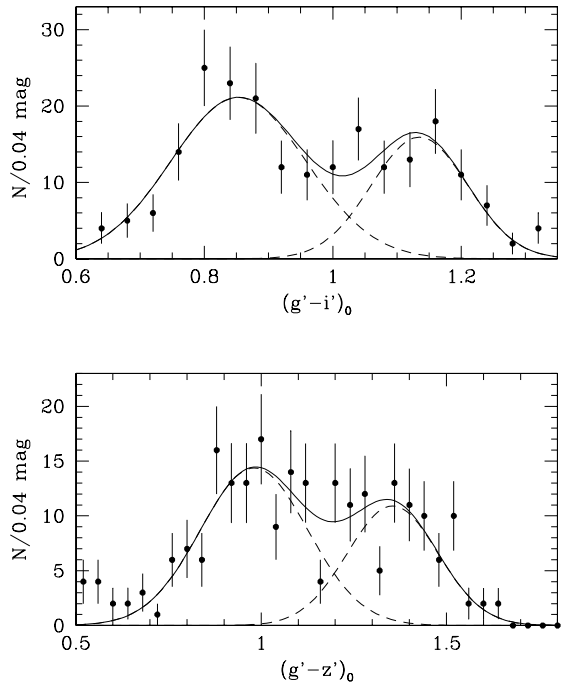}
\caption[]{\emph{Upper panel:} Histogram of $(g'-i')_0$ colors for
our sample of GCs in NGC 5128.  The dashed
lines give the bimodal deconvolution
for the color distribution described in the text, while the solid line is the sum
of both components.
\emph{Lower panel:} Histogram of $(g'-z')_0$ colors for the same
clusters.  As above, the dashed lines give the bimodal
solution for the color distribution, while the solid line is the 
total for the bimodal fit.}
\label{fig:cfit}
\end{figure}

To estimate the mean colors of each mode, we used the magnitude interval $18 < r' < 20$
to avoid excessive random errors at the faint end and any systematic effects 
of the mass/metallicity relation at the bright end (see Harris 2009a\nocite{harris09a}). 
We used RMIX\footnote{RMIX is publicly available at http://www.math.mcmaster.ca/peter/mix/mix.html} 
to fit bimodal Gaussian distributions, yielding 
$\langle g'-i' \rangle_0 = 0.85 \pm 0.02$ (blue), $1.13 \pm 0.02$ (red), and also 
$\langle g'-z' \rangle_0 = 0.98 \pm 0.03$ (blue), $1.35 \pm 0.04$ (red).
The relative fractions of GCs are $0.63 \pm 0.07$ in the blue and $0.37 \pm 0.07$ in the red.
It is worth noting here that
\citet{woodley10a} found tentative evidence for a trimodal metallicity distribution
directly from the spectroscopic line indices for about 70 clusters.  The third,
intermediate-color mode may be due to a small
proportion of somewhat younger clusters \cite[see][for additional discussion]{woodley10a}.
We find no clear evidence for a third, intermediate mode in our data, but larger and more 
precise samples may reveal it. 
The color histograms from our data, along with the bimodal 
solution described above, are shown in Figure \ref{fig:cfit}.

\begin{figure*}[hbt!]
\epsscale{1.0}
\centering8
\plotone{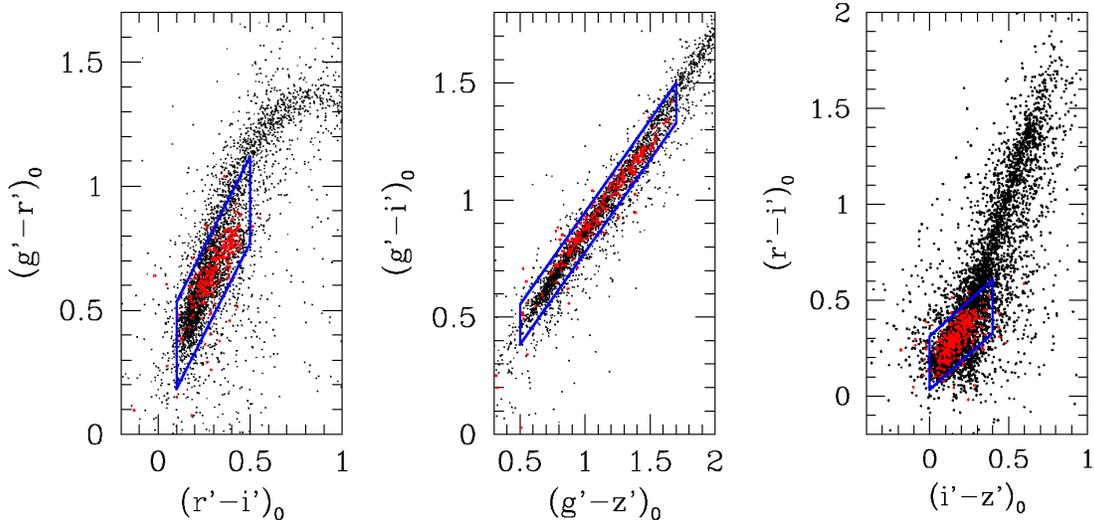}
\caption[]{Color-color diagrams for the NGC~5128 field objects, with the known GCs over-plotted in red. The blue boxes highlight regions encapsulating $90\%$ of the GCs in each of the three color-color planes.}
\label{fig:ccdsall}
\end{figure*}

This NGC~5128 data can now be used to investigate the 
intrinsic colors of GCs in the $g'r'i'z'$ indices. 
As already established in previous photometric and spectroscopic work 
\citep[e.g.][]{woodley10a,harris04a}, the great majority of the GCs in this 
galaxy follow the classic pattern of large age and bimodal metallicity distribution 
like other giant ellipticals, so their range in color is already well restricted, with 
few genuinely young, blue objects. 
To specify these intrinsic colors a bit further, we look for their distributions in color-color space, 
relative to the field stars. 
Figure~\ref{fig:ccdsall} shows three of the possible six color-color diagrams,
again including all measured NGC~5128 sources, with the GCs over-plotted in red. 
In each graph, we define regions containing $90\%$ of the GC population.  Objects outside these
boxes have a high probability of being either contaminants or star clusters
that are not classically old. 
The slopes of the boxes shown in Figure~\ref{fig:ccdsall} are 
$\Delta(g'-r') / \Delta(r'-i')$ = 1.47, 
$\Delta(g'-i') / \Delta(g'-z')$ = 0.79, and 
$\Delta(r'-i') / \Delta(i'-z')$ = 0.74. 
The regions of higher GC density highlight the \emph{intrinsic colors of typical GCs} 
in these colors, showing that the typical GC population is confined to small regions in color-color 
space and that half of the field objects can be rejected this way.  
Perhaps the most effective single diagrams are $(g'-r')$ versus $(r'-i')$ and $(r'-i')$ versus $(i'-z')$ which define
the narrowest zones.

\section{Calibration versus Metallicity }
\label{sec:calib}
As mentioned above, one of the goals of our study was to define the baseline conversion of the 
$g'r'i'z'$ color indices to metallicity, for ``typical'' old GCs.  Ideally, transformation of a given color index 
to (say) [Fe/H] would be done by having in hand both the photometric indices and the [Fe/H] values as determined by high-dispersion spectroscopy. 
\citet{peng06} present a preliminary calibration of this type for $(g-z)$ versus [Fe/H], using their unpublished photometry for 40 Milky Way clusters, plus 55 more from the Virgo giants M87 and M49 (see their Figs. 11 and 12), along with a variety of literature sources for the spectroscopy.

In our case, \emph{high-dispersion} and high-signal-to-noise ($S/N$) spectroscopy for the 
clusters in NGC~5128 are as yet available for only a small number 
of objects \citep[see, e.g.][]{rejkuba07,taylor10}.
However, \citet{woodley10a} present a recent 
spectroscopic study of a large sample of the NGC~5128 and Milky Way GCs 
through the use of Lick indices.  A significant advantage of their database 
is that it analyzes the line indices of clusters within both galaxies on 
an internally homogeneous and self-consistent system, allowing a more reliable comparison.  
We use the material from their study here, realizing that it may 
be superseded once a more extensive database of high-dispersion spectroscopic metallicities becomes available.

\begin{figure}[hbt!]
\epsscale{1.0}
\centering
\plotone{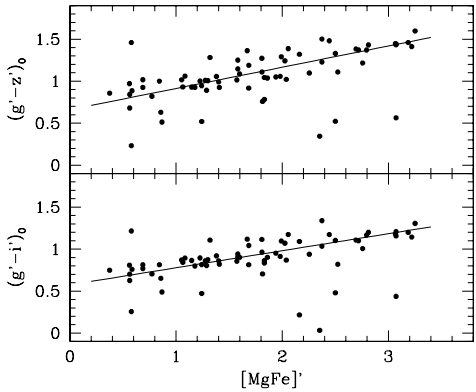}
\caption[]{Correlation of intrinsic color indices $(g'-i')_0$ and $(g'-z')_0$ against the
Lick metallicity index [MgFe]$'$ for the GCs in NGC~5128, with spectroscopic data from \citet{woodley10a}.}
\label{mgfe_color}
\end{figure}

Here we work with the two most metallicity-sensitive 
color indices ($g'-i'$, $g'-z'$) and derive their transformations into 
[Fe/H] in two steps.  The first of these is their correlation against 
the Lick index [MgFe]$'$, defined as 
$${\rm [MgFe]}' \, = \, \sqrt{Mg_b \times (0.72 Fe5270 \, + \, 0.28 Fe5335) } $$ \citep{thomas03}.  
[MgFe]$'$ is designed to be a relatively clean heavy-element abundance indicator, 
highly insensitive to [$\alpha$/Fe] variations.  In 
Figure \ref{mgfe_color}, we show the dereddened color indices versus this index 
for the NGC~5128 clusters in our catalog that are
in common with \citet{woodley10a}.  We find in both cases that linear 
relations match the data well, given by
\begin{equation}
\begin{split}
\centering
(g'-i')_0 = & (0.575 \pm 0.039) + (0.203 \pm 0.021) {\rm [MgFe]}' \\ 
	& (n=61)
\end{split}
\end{equation}
\begin{equation}
\begin{split}
\centering
(g'-z')_0 = & (0.660 \pm 0.048) + (0.253 \pm 0.025) {\rm [MgFe]}' \\
           & (n=59).
\end{split}
\end{equation}

In both cases the rms scatter around these mean lines is $\pm 0.13$ mag in color.

\begin{figure}[hbt!]
\epsscale{1.0}
\centering
\plotone{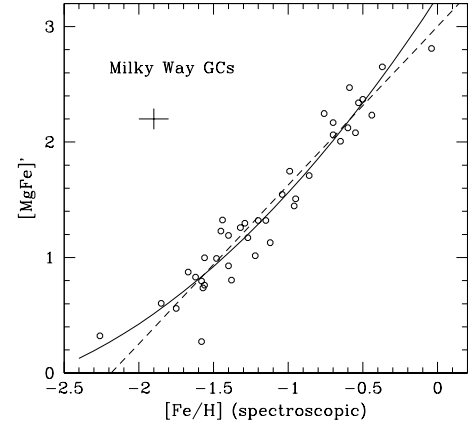}
\caption[]{Correlation of the Lick metallicity index [MgFe]$'$ against heavy-element abundance
[Fe/H] for GCs in the Milky Way.  The dashed and solid lines show the linear
and quadratic best-fit solutions given in the text. A typical 
error bar is shown as a cross in the upper left corner.}
\label{MWcal}
\end{figure}

The second step is to transform [MgFe]$'$ to metallicity, [Fe/H].  Here we 
use the Milky Way GCs, which have metallicity measurements
superior to those of any other galaxy.
We adopt [Fe/H] values from the catalog of \citet{harris96} (2003 edition) 
and correlate these against their [MgFe]$'$ values measured in \citet{woodley10a} from Milky Way GC spectra 
obtained from \citet{puzia02} and \citet{schiavon05}.  
The results for 40 Milky Way clusters are shown in Figure \ref{MWcal}.  
We note here that the catalog values of [Fe/H] were originally based on the \citet{zinn84} scale.  
However, a large number of abundance measurements from high-dispersion spectroscopy have been 
added, making the current Milky Way catalog list of metallicities 
close to the \citet{carretta97} scale.

An interesting question raised by Figure~\ref{MWcal} is whether or not 
the conversion is adequately matched by a linear relation.  
From \citet{puzia02}, \citet{woodley10a}, and \citet{harris96}, we estimate that the typical 
uncertainty in each quantity is $\pm0.1$~dex. 
We find the best-fit linear or quadratic relations to be
\begin{equation}
\centering
{\rm [MgFe]}' = (2.987 \pm 0.075) + (1.347 \pm 0.062) {\rm [Fe/H]} \hspace{0.3cm }{\rm (linear)}
\end{equation}
\vspace{0.1cm}
\begin{equation}
\begin{split}
\centering
{\rm [MgFe]}' =&  (3.253\pm 0.120) + (1.966 \pm 0.230) {\rm [Fe/H]} \\
               & + (0.281 \pm 0.104) {\rm [Fe/H]}^2 \hspace{0.3cm} {\rm (quadratic)}.
\end{split}
\end{equation}

The rms scatter around these relations is $\pm 0.19$ (linear) and $\pm 0.17$ (quadratic).  
Given also that the solution for the [Fe/H]$^2$ term is significant at 
the $3 \sigma$ level, we therefore find a slight preference for the quadratic solution and recommend it for future use.
We note that \citet{peng06} also found that a single linear transformation for $(g-z)$ to [Fe/H] was not suitable.
Fig.~\ref{MWcal} shows, however, that in the range $-1.7 \lesssim$ [Fe/H] $\lesssim -0.3$ 
that contains the great majority of clusters, the difference between the linear and quadratic conversions
is small.

Combining the two steps outlined above, our color to metallicity conversions are
\begin{equation}
\centering
(g'-i')_0  = 1.235  +  0.399 {\rm [Fe/H]} + 0.057 {\rm [Fe/H]}^2,
\end{equation}
\begin{equation}
\centering
(g'-z')_0  = 1.483 + 0.497 {\rm [Fe/H]} + 0.071 {\rm [Fe/H]}^2,
\end{equation}

At a mean typical GC metallicity [Fe/H] $\simeq -1$, we 
therefore have $\Delta (g'-i') / \Delta [Fe/H] \simeq 0.29$ 
and $\Delta (g'-z') / \Delta [Fe/H] \simeq 0.36$.  With good photometric 
measurement precisions of $\pm 0.05$, the $g'i'z'$ indices can then be 
used to predict metallicities to within internal uncertainties of 
typically $\pm 0.15$ dex.  

\begin{figure}[hbt!]
\epsscale{1.2}
\centering
\plotone{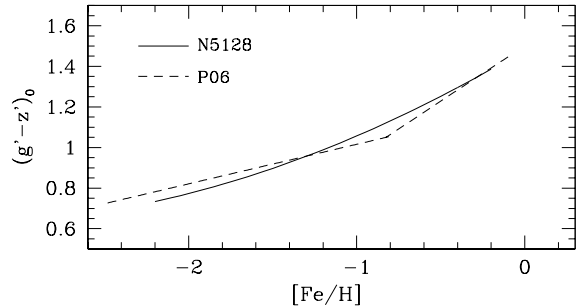}
\caption[]{Comparison of our calibration of $(g'-z')_0$ versus metallicity
(solid line) with the \citet{peng06} calibration (dashed line). While the \citet{peng06} 
calibration is carried out on the F475W and F850W ($g$ and $z$) Advanced Camera for Surveys (ACS) filters, we plot their 
transformation here as $(g'-z')_0$ since the differences between the filter systems are 
negligible for the purposes of this comparison.
}
\label{fehcomp}
\end{figure}

We can compare our recommended conversion of $(g'-z')_0$ to [Fe/H]
with that of \citet{peng06}; both curves are shown in Figure \ref{fehcomp}. 
For this purpose, the very small differences between the $gz$ and $g'z'$ systems 
mentioned earlier are unimportant. 
Peng et al.~adopt a two-section linear curve with the changeover point
at [Fe/H] $\sim -1$, roughly halfway between the normal metal-poor
and metal-rich GC sequences, and compare their data with several 
population-synthesis models (see their Figure 12).  
The internal scatter of their [Fe/H] data points around their
adopted calibration is not specifically listed but appears to be $\pm0.2-0.3$ dex.
The differences between their curve and the one we have derived here
are within $\sim 0.1$ dex over most of the range.
While the Milky Way GCs constrain our transformation in the range 
-2.0 $\lesssim$ [Fe/H] $\lesssim$ -0.5 (0.7 $\lesssim$ $(g'-z')_0$ $\lesssim$ 1.3), 
we note our transformation agrees well with Peng et al.~for [Fe/H] $\gtrsim$ -0.5.
Close inspection of the data used by 
Peng et al.~(see their Fig.~12) indicates
that our continuous quadratic transformation falls close to
the centroid of their data, and fits the calibration
as well as any of the various model curves shown there. 

The conversions derived above can be used to determine the mean 
metallicities of the NGC~5128 GCs.  If we adopt the mean colors obtained in
Section~\ref{sec:diagrams}, we find [Fe/H] = $-1.19$ (blue), $-0.27$ (red) 
from a direct average of $(g'-i')_0$ and $(g'-z')_0$.  
The two colors give the same metallicities to within $\pm 0.05$ dex.  
In addition, the \citet{peng06} calibration
for $(g-z)$ gives the same means to within $\pm 0.04$ dex.  
These values are $\simeq 0.1$ dex more metal-rich than has often
been obtained in other giant ellipticals \citep[e.g.][]{geisler96,brodie06,harris09a},
but they fall within the expected range given the combined internal uncertainties
in the photometric calibration, reddenings, and transformation coefficients.

Another consistency check of the external
accuracy of these transformations can be done by comparing our results with 
the $gri$ photometry of the M87 cluster system by \citet{harris09b}.  The 
data from that study consist of photometry for several thousand GCs throughout the 
M87 halo, placed accurately on the SDSS standard system by reference to 
the SDSS-DR5 catalog of sources in the same region.  The 
blue and red GC sequences in M87 are located at mean intrinsic colors 
$\langle g-i \rangle_0$(blue) = 0.80, $\langle g-i \rangle_0$(red) = 1.07, obtained using $E_{g-i} = 0.03$.  
These translate through the preceding equations into mean metallicities 
of [Fe/H] = $-1.47$ (blue), $-0.56$ (red), which are both within 
$\pm 0.1$ dex of the metallicities for the two sequences 
in giant galaxies obtained through a variety of other methods \citep[e.g.][]{brodie06}.

\section{Comparison of Filter Systems}
\label{sec:filtersystems}
As noted in the Introduction, broadband photometry with a metallicity-sensitive
color index is a fast and effective way to derive first-order MDFs
for GC systems, though ultimately not as accurate
(or internally precise) as spectroscopic indices.  Many different color
indices have been used in published studies for this purpose, and it
is of interest to gauge how the $u'g'r'i'z'$ filter system compares with others. 
The well known Johnson/Cousins
$(V-I)$ is not an especially sensitive index \citep[see][]{barmby00}, but it
was adopted particularly in many early HST-based studies
that were crucial to defining the bimodality paradigm
\citep[e.g.][among many others]{gebhardt99,larsen01} essentially because the heavily
used Wide Field Planetary Camera 2 (WFPC2) camera had relatively low blue response. 
Two other common indices 
are $(B-I)$ \citep[e.g.][]{harris09a}
and $(C-T_1)$ \citep[e.g.][]{geisler96,harris04b}.  For comparison, the equations
linking them to [Fe/H] are
\begin{equation}
(V-I)_0 \, = \, 1.15 + 0.156 {\rm [Fe/H]}
\end{equation}
\begin{equation}
(B-I)_0 \, = \, 2.158 + 0.375 {\rm [Fe/H]}
\end{equation}
\begin{equation}
(C-T_1)_0 \, = \, 1.998 +  0.748 {\rm [Fe/H]} + 0.138 {\rm [Fe/H]}^2
\end{equation}
trim whitespace on eps
Their slopes near the mid-range [Fe/H] $\simeq -1$ are 0.156 for $(V-I)$,
0.37 for $(B-I)$, and 0.47 for $(C-T_1)$, compared to
0.285 for $(g'-i')$ and 0.355 for $(g'-z')$.  
Of these five, the Washington index $(C-T_1)_0$ appears to be the most sensitive to metallicity because 
of its wide baseline and also because the $C$ filter 
is positioned directly over a large number of heavy-element absorption lines.  
A more complete comparison of the Johnson/Cousins color indices,
including the various pairs that can be constructed from $UBVRIJK$, is given by \citet{barmby00}.

The use of the near-UV filter in the $u'g'r'i'z'$ system could in principle
yield an even more sensitive index such as $(u'-i')$ or $(u'-z')$,
but (for most ground-based cameras) at a huge penalty in exposure
time.  Combinations of optical filters with near-infrared filters have also been explored, such as the combination of
$B$ and $R$ with the 3.6$\mu$ color 
\citep{spitler08}, or $(V-K)$ \citep{barmby00}.  These have a very wide baseline and can ``split''
the bimodal GC sequences more clearly, but have the disadvantage that the observations
require two sets of instrumentation.

We regard both $(g'-i')_0$ and $(g'-z')_0$ as competitive color indices
for GC MDF photometry, though both $(B-I)_0$ and $(C-T_1)_0$ 
continue to be effective for this purpose as well.  
These four indices all provide good compromises between intrinsic metallicity
sensitivity and the necessary exposure times to obtain precise photometry.
For ground-based photometry, an extra factor to consider is the 
degree of fringing present in the red or
near-infrared filters.  For example, the fringing was fairly large in the $z'$ filter in this work
(as noted above) and this may reduce the internal precision that 
can be achieved for $(g'-z')$, offsetting the advantage of its
wider color baseline. 

\begin{figure}[hbt!]
\epsscale{0.8}
\centering
\plotone{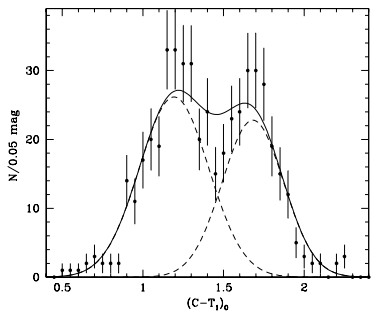}
\caption[]{Color distribution in the Washington index $(C-T_1)_0$ for 500 clusters
in NGC 5128.  The bimodal fit to the data is shown in the dashed lines for the
blue and red modes, with the sum in the solid line.
}
\label{ct1fit}
\end{figure}

The data in the \citet{woodley10b} catalog also provide a convenient summary of the
$(C-T_1)$ color indices extracted from the Washington-system survey of the
NGC 5128 field \citep{harris04a}.  The color distribution for 500 clusters in
the magnitude range $17 < T_1 < 21$ is shown in Figure \ref{ct1fit}.
Earlier versions of the $(C-T_1)$ distribution based on much smaller numbers of
individually selected clusters were derived in \citet{harris92,woodley05,woodley10a}.  
A bimodal-Gaussian fit to this distribution yields mean colors for the two
modes of $1.413 \pm 0.038$ (blue), $1.902 \pm 0.038$ (red), with proportions
$0.565 \pm 0.074$ (blue), $0.435 \pm 0.074$ (red).  There is no significant presence
of additional modes.  


\section{Conclusions}
\label{sec:conclusions}
In this study we have used new ground-based photometry to generate
a database of $g'r'i'z'$ indices for the
GCs in NGC~5128, the nearest giant elliptical galaxy.
Our data have been calibrated against fundamental USNO standard
stars and thus provide among the first comprehensive calibrations
of intrinsic GC colors in the $g'r'i'z'$ system.
Our final data list contains colors and magnitudes for 323
known GCs covering the metallicity range $-2 <$ [Fe/H] $< 0$.

Adding this material to previously published high-S/N spectroscopic
indices for clusters in NGC~5128 and the Milky Way, we derive 
transformations between $(g'-i')_0, (g'-z')_0$, [MgFe]$'$, and [Fe/H].
These transformations can be used to determine metallicity
distributions from the USNO color indices to a typical precision
of $\pm0.15$ dex.
Comparison with other widely used photometric indices indicates
that the colors constructed from $g'i'z'$ are competitive with
other broadband optical indices such as $(B-I)$ or $(C-T_1)$.

\bibliographystyle{apj}
\bibliography{sinnott2010_corrected}

\clearpage
\appendix
\begin{longtable}{cccccccccccc}
\centering
\tabletypesize{\scriptsize} 
\tablecaption{ $g'r'i'z'$ Photometry Catalog Of NGC~5128 GCs
\label{tbl:GCgriz}}
\tablehead{ 
\colhead{ID} & \colhead{$\alpha$} & \colhead{$\delta$} & \colhead{V} & \colhead{$g'$} &  \colhead{$\sigma_{g'}$} & \colhead{$r'$} &  \colhead{$\sigma_{r'}$} & \colhead{$i'$} &  \colhead{$\sigma_{i'}$} & \colhead{$z'$} &  \colhead{$\sigma_{z'}$}\\ \colhead{} & \colhead{(J2000)} & \colhead{(J2000)} &
} 
GC001 & 00:53:40.08 & -42:56:51.995 & 18.75 & 19.255 & 0.033  & 18.477 & 0.025  & 18.086 & 0.034  & 17.858 & 0.031 \\
GC039 & 00:53:38.58 & -43:06:26.660 & 19.45 & 20.046 & 0.059  & 19.129 & 0.034  & 18.683 & 0.038  & 18.417 & 0.037 \\
GC043 & 00:53:38.70 & -42:53:35.130 & 99.00 & 99.000 & 99.00  & 21.718 & 0.068  & 21.188 & 0.091  & 99.000 & 99.00 \\
GC045 & 00:53:38.74 & -42:59:48.380 & 20.23 & 20.439 & 0.086  & 19.891 & 0.047  & 19.442 & 0.048  & 19.137 & 0.046 \\
GC046 & 00:53:38.75 & -43:01:45.605 & 19.22 & 19.708 & 0.054  & 19.017 & 0.038  & 18.722 & 0.042  & 18.552 & 0.039 \\
GC047 & 00:53:38.90 & -43:08:43.215 & 18.80 & 19.184 & 0.044  & 18.538 & 0.033  & 18.298 & 0.038  & 18.166 & 0.036 \\
GC048 & 00:53:38.91 & -42:53:07.105 & 19.85 & 20.433 & 0.068  & 19.577 & 0.025  & 19.068 & 0.035  & 18.734 & 0.032 \\
GC049 & 00:53:38.91 & -42:58:16.350 & 19.51 & 19.966 & 0.056  & 19.276 & 0.034  & 18.935 & 0.038  & 18.765 & 0.038 \\
GC052 & 00:53:39.02 & -42:59:33.490 & 19.30 & 19.553 & 0.048  & 19.087 & 0.034  & 18.894 & 0.038  & 18.830 & 0.038 \\
GC053 & 00:53:39.05 & -43:02:24.460 & 19.89 & 20.531 & 0.076  & 19.606 & 0.035  & 19.116 & 0.040  & 18.785 & 0.039 \\
GC054 & 00:53:39.10 & -43:04:11.555 & 19.16 & 19.687 & 0.054  & 99.000 & 0.037  & 18.677 & 0.041  & 18.528 & 0.042 \\
GC055 & 00:53:39.11 & -43:01:18.385 & 18.95 & 19.459 & 0.049  & 18.758 & 0.036  & 18.431 & 0.040  & 18.275 & 0.039 \\
GC056 & 00:53:39.14 & -43:06:01.640 & 18.91 & 19.581 & 0.050  & 18.605 & 0.035  & 18.097 & 0.038  & 17.756 & 0.037 \\
GC057 & 00:53:39.16 & -42:58:29.825 & 19.10 & 19.591 & 0.048  & 18.850 & 0.034  & 18.501 & 0.038  & 18.325 & 0.034 \\
GC058 & 00:53:39.16 & -42:57:51.130 & 19.73 & 20.289 & 0.066  & 19.429 & 0.034  & 18.951 & 0.038  & 18.649 & 0.038 \\
GC060 & 00:53:39.20 & -43:08:14.200 & 18.49 & 19.058 & 0.046  & 18.229 & 0.038  & 17.856 & 0.041  & 17.675 & 0.040 \\
GC061 & 00:53:39.25 & -42:52:35.325 & 99.00 & 20.431 & 0.069  & 19.733 & 0.027  & 19.353 & 0.037  & 19.213 & 0.042 \\
GC062 & 00:53:39.26 & -42:57:48.435 & 19.05 & 19.555 & 0.050  & 18.853 & 0.036  & 18.515 & 0.040  & 18.328 & 0.038 \\
GC063 & 00:53:39.29 & -43:08:17.650 & 20.34 & 20.845 & 0.096  & 20.134 & 0.038  & 19.788 & 0.043  & 19.550 & 0.052 \\
GC064 & 00:53:39.34 & -43:07:36.160 & 20.44 & 20.818 & 0.094  & 20.366 & 0.037  & 20.050 & 0.043  & 19.820 & 0.055 \\
GC066 & 00:53:39.39 & -43:01:22.860 & 18.86 & 19.466 & 0.049  & 18.580 & 0.036  & 18.135 & 0.039  & 17.872 & 0.038 \\
GC068 & 00:53:39.47 & -43:04:32.645 & 20.08 & 99.000 & 0.065  & 19.384 & 0.047  & 19.119 & 0.053  & 19.026 & 0.072 \\
GC070 & 00:53:39.55 & -43:04:34.690 & 19.20 & 19.702 & 0.051  & 18.965 & 0.035  & 18.637 & 0.038  & 18.490 & 0.039 \\
GC071 & 00:53:39.60 & -43:04:24.460 & 19.92 & 20.369 & 0.068  & 19.669 & 0.035  & 19.388 & 0.038  & 19.268 & 0.044 \\
GC072 & 00:53:39.61 & -42:54:50.440 & 19.25 & 19.714 & 0.041  & 19.051 & 0.028  & 18.735 & 0.035  & 18.568 & 0.034 \\
GC073 & 00:53:39.62 & -43:03:15.415 & 19.84 & 20.265 & 0.067  & 19.692 & 0.036  & 19.350 & 0.040  & 19.250 & 0.042 \\
GC074 & 00:53:39.62 & -42:53:24.568 & 17.25 & 17.807 & 0.033  & 16.936 & 0.026  & 16.497 & 0.035  & 16.235 & 0.032 \\
GC075 & 00:53:39.63 & -43:05:34.615 & 19.96 & 20.553 & 0.078  & 19.645 & 0.034  & 19.180 & 0.039  & 18.835 & 0.039 \\
GC076 & 00:53:39.64 & -42:48:59.070 & 19.44 & 19.926 & 0.068  & 19.181 & 0.039  & 18.847 & 0.091  & 18.719 & 0.084 \\
GC077 & 00:53:39.65 & -43:01:21.680 & 17.91 & 18.556 & 0.043  & 17.640 & 0.039  & 17.187 & 0.042  & 16.914 & 0.040 \\
GC079 & 00:53:39.70 & -43:09:58.330 & 99.00 & 99.000 & 99.00  & 22.132 & 0.097  & 21.480 & 0.108  & 99.000 & 99.00 \\
GC082 & 00:53:39.74 & -42:54:29.540 & 99.00 & 20.412 & 0.055  & 19.529 & 0.026  & 19.026 & 0.035  & 18.705 & 0.032 \\
GC083 & 00:53:39.74 & -43:10:16.430 & 20.45 & 20.946 & 0.146  & 20.251 & 0.044  & 20.144 & 0.052  & 20.196 & 0.070 \\
GC085 & 00:53:39.81 & -43:08:42.590 & 99.00 & 99.000 & 99.00  & 21.041 & 0.058  & 20.492 & 0.058  & 20.102 & 0.071 \\
GC086 & 00:53:39.83 & -43:01:08.095 & 18.38 & 18.963 & 0.042  & 18.057 & 0.034  & 17.594 & 0.038  & 17.293 & 0.035 \\
GC087 & 00:53:39.83 & -42:59:23.305 & 18.74 & 19.250 & 0.046  & 18.545 & 0.035  & 18.221 & 0.038  & 18.031 & 0.036 \\
GC088 & 00:53:39.84 & -43:05:32.685 & 20.02 & 20.657 & 0.087  & 19.814 & 0.041  & 19.313 & 0.046  & 19.059 & 0.051 \\
GC089 & 00:53:39.85 & -42:55:48.370 & 99.00 & 21.028 & 0.107  & 20.841 & 0.040  & 20.911 & 0.064  & 99.000 & 99.00 \\
GC090 & 00:53:39.86 & -42:52:04.920 & 99.00 & 99.000 & 99.00  & 21.787 & 0.071  & 21.249 & 0.082  & 99.000 & 99.00 \\
GC091 & 00:53:39.88 & -42:56:10.305 & 17.71 & 18.314 & 0.032  & 17.439 & 0.028  & 16.989 & 0.037  & 16.714 & 0.035 \\
GC095 & 00:53:39.99 & -43:09:08.480 & 99.00 & 21.202 & 0.183  & 20.474 & 0.038  & 20.216 & 0.048  & 20.009 & 0.058 \\
GC096 & 00:53:40.01 & -42:54:08.835 & 20.43 & 20.953 & 0.080  & 20.146 & 0.027  & 19.735 & 0.037  & 19.502 & 0.044 \\
GC098 & 00:53:40.04 & -43:05:30.190 & 19.74 & 20.314 & 0.068  & 19.475 & 0.035  & 19.048 & 0.039  & 18.813 & 0.036 \\
GC102 & 00:53:40.10 & -43:08:32.910 & 99.00 & 99.000 & 99.00  & 21.290 & 0.050  & 20.732 & 0.060  & 99.000 & 99.00 \\
GC103 & 00:53:40.11 & -42:54:40.783 & 19.37 & 19.602 & 0.039  & 19.362 & 0.028  & 19.201 & 0.037  & 19.083 & 0.037 \\
GC104 & 00:53:40.11 & -42:50:51.150 & 99.00 & 99.000 & 99.00  & 21.093 & 0.031  & 20.646 & 0.052  & 20.619 & 0.089 \\
GC106 & 00:53:40.12 & -43:09:25.375 & 18.04 & 18.507 & 0.042  & 17.767 & 0.037  & 17.482 & 0.040  & 17.323 & 0.039 \\
GC107 & 00:53:40.12 & -42:52:27.597 & 99.00 & 20.774 & 0.087  & 20.075 & 0.026  & 19.747 & 0.038  & 19.550 & 0.041 \\
GC109 & 00:53:40.21 & -42:56:25.345 & 18.93 & 19.482 & 0.037  & 18.630 & 0.026  & 18.193 & 0.035  & 17.909 & 0.032 \\
GC110 & 00:53:40.21 & -43:03:02.415 & 18.68 & 19.231 & 0.044  & 18.387 & 0.034  & 17.954 & 0.038  & 17.685 & 0.035 \\
GC111 & 00:53:40.22 & -42:57:40.555 & 19.73 & 20.186 & 0.062  & 19.466 & 0.034  & 19.115 & 0.038  & 18.891 & 0.040 \\
GC113 & 00:53:40.22 & -42:50:45.932 & 18.44 & 19.036 & 0.037  & 18.105 & 0.026  & 17.604 & 0.035  & 17.291 & 0.033 \\
GC115 & 00:53:40.25 & -42:51:21.545 & 99.00 & 21.158 & 0.118  & 20.473 & 0.028  & 20.237 & 0.042  & 20.080 & 0.081 \\
GC116 & 00:53:40.28 & -43:00:19.630 & 19.34 & 19.744 & 0.050  & 19.053 & 0.033  & 18.846 & 0.038  & 18.632 & 0.035 \\
GC118 & 00:53:40.31 & -43:07:21.705 & 19.61 & 20.044 & 0.059  & 19.380 & 0.033  & 19.103 & 0.038  & 18.964 & 0.037 \\
GC119 & 00:53:40.32 & -43:09:38.790 & 18.60 & 19.064 & 0.044  & 18.305 & 0.035  & 18.009 & 0.039  & 17.827 & 0.036 \\
GC120 & 00:53:40.33 & -42:57:15.285 & 17.41 & 17.967 & 0.038  & 17.115 & 0.034  & 16.694 & 0.038  & 16.427 & 0.034 \\
GC121 & 00:53:40.35 & -42:58:05.780 & 19.62 & 20.182 & 0.062  & 19.340 & 0.034  & 18.934 & 0.038  & 18.685 & 0.039 \\
GC126 & 00:53:40.49 & -43:08:29.535 & 20.78 & 99.000 & 99.00  & 20.563 & 0.041  & 20.207 & 0.055  & 20.017 & 0.080 \\
GC127 & 00:53:40.49 & -43:06:20.455 & 20.04 & 20.580 & 0.078  & 19.833 & 0.035  & 19.471 & 0.042  & 19.273 & 0.045 \\
GC129 & 00:53:40.51 & -43:01:15.190 & 18.12 & 18.708 & 0.041  & 17.815 & 0.034  & 17.363 & 0.038  & 17.062 & 0.035 \\
GC130 & 00:53:40.57 & -43:02:57.155 & 18.91 & 19.553 & 0.051  & 18.679 & 0.043  & 18.215 & 0.041  & 17.925 & 0.041 \\
GC131 & 00:53:40.59 & -43:09:09.475 & 19.42 & 19.882 & 0.075  & 19.181 & 0.035  & 18.879 & 0.040  & 18.732 & 0.039 \\
GC132 & 00:53:40.59 & -43:04:14.795 & 19.29 & 19.902 & 0.055  & 19.048 & 0.035  & 18.632 & 0.039  & 18.320 & 0.036 \\
GC135 & 00:53:40.63 & -42:55:18.493 & 19.12 & 19.560 & 0.037  & 18.898 & 0.026  & 18.579 & 0.035  & 18.411 & 0.032 \\
GC136 & 00:53:40.68 & -43:02:06.670 & 19.71 & 20.374 & 0.069  & 19.356 & 0.033  & 18.876 & 0.038  & 18.531 & 0.035 \\
GC137 & 00:53:40.68 & -42:55:09.492 & 19.46 & 19.972 & 0.043  & 19.119 & 0.025  & 18.655 & 0.035  & 18.321 & 0.033 \\
GC138 & 00:53:40.68 & -42:53:32.802 & 17.80 & 18.405 & 0.031  & 17.556 & 0.026  & 17.102 & 0.035  & 16.837 & 0.033 \\
GC140 & 00:53:40.70 & -43:03:23.870 & 20.40 & 99.000 & 0.107  & 20.162 & 0.038  & 19.731 & 0.045  & 19.407 & 0.054 \\
GC141 & 00:53:40.74 & -43:01:31.905 & 99.00 & 20.441 & 0.072  & 19.522 & 0.035  & 19.031 & 0.039  & 18.704 & 0.037 \\
GC142 & 00:53:40.74 & -42:58:03.165 & 18.92 & 19.309 & 0.043  & 18.682 & 0.032  & 18.442 & 0.037  & 18.315 & 0.034 \\
GC143 & 00:53:40.74 & -43:03:09.470 & 19.51 & 99.000 & 0.057  & 19.343 & 0.035  & 18.982 & 0.040  & 18.821 & 0.037 \\
GC144 & 00:53:40.80 & -43:04:19.145 & 19.78 & 20.403 & 0.070  & 19.535 & 0.035  & 19.116 & 0.039  & 18.847 & 0.039 \\
GC145 & 00:53:40.81 & -42:57:25.280 & 18.62 & 19.162 & 0.045  & 18.372 & 0.035  & 18.002 & 0.038  & 17.811 & 0.037 \\
GC147 & 00:53:40.82 & -42:58:07.690 & 20.44 & 20.983 & 0.104  & 20.188 & 0.035  & 19.818 & 0.043  & 19.587 & 0.050 \\
GC149 & 00:53:40.86 & -42:57:00.157 & 19.57 & 20.019 & 0.046  & 19.367 & 0.026  & 19.035 & 0.035  & 18.869 & 0.037 \\
GC150 & 00:53:40.86 & -43:07:59.020 & 18.20 & 18.725 & 0.040  & 17.888 & 0.033  & 17.520 & 0.038  & 17.286 & 0.034 \\
GC152 & 00:53:40.88 & -43:02:31.230 & 99.00 & 99.000 & 99.00  & 21.189 & 0.052  & 20.803 & 0.070  & 99.000 & 99.00 \\
GC153 & 00:53:40.89 & -42:52:12.122 & 19.82 & 20.562 & 0.074  & 19.725 & 0.036  & 19.276 & 0.047  & 19.013 & 0.049 \\
GC154 & 00:53:40.92 & -43:07:32.380 & 20.70 & 21.223 & 0.189  & 20.417 & 0.037  & 20.033 & 0.045  & 19.686 & 0.052 \\
GC155 & 00:53:40.93 & -42:57:42.635 & 18.72 & 19.036 & 0.041  & 18.514 & 0.032  & 18.285 & 0.037  & 18.170 & 0.033 \\
GC158 & 00:53:40.95 & -43:07:23.395 & 19.55 & 20.177 & 0.063  & 19.281 & 0.037  & 18.799 & 0.040  & 18.523 & 0.039 \\
GC160 & 00:53:41.01 & -42:50:30.400 & 19.87 & 20.395 & 0.086  & 19.676 & 0.040  & 19.409 & 0.091  & 19.293 & 0.086 \\
GC161 & 00:53:41.01 & -42:57:45.665 & 20.25 & 20.696 & 0.085  & 20.053 & 0.035  & 19.764 & 0.043  & 19.620 & 0.047 \\
GC162 & 00:53:41.02 & -43:08:39.105 & 20.78 & 99.000 & 99.00  & 20.477 & 0.036  & 20.112 & 0.045  & 19.774 & 0.052 \\
GC164 & 00:53:41.06 & -43:06:03.140 & 20.39 & 20.957 & 0.101  & 20.154 & 0.038  & 19.765 & 0.042  & 19.564 & 0.047 \\
GC165 & 00:53:41.07 & -43:05:06.320 & 19.59 & 20.103 & 0.062  & 19.371 & 0.036  & 19.031 & 0.041  & 18.807 & 0.039 \\
GC170 & 00:53:41.10 & -43:03:32.985 & 20.44 & 21.051 & 0.113  & 20.128 & 0.035  & 19.802 & 0.043  & 19.648 & 0.048 \\
GC173 & 00:53:41.13 & -43:03:07.880 & 99.00 & 20.921 & 0.151  & 20.336 & 0.038  & 99.000 & 0.049  & 19.881 & 0.061 \\
GC174 & 00:53:41.13 & -43:09:27.900 & 19.43 & 19.858 & 0.073  & 19.160 & 0.035  & 18.833 & 0.040  & 18.651 & 0.040 \\
GC180 & 00:53:41.22 & -42:53:04.598 & 19.07 & 19.607 & 0.039  & 18.839 & 0.026  & 18.467 & 0.035  & 18.261 & 0.034 \\
GC181 & 00:53:41.23 & -43:04:09.685 & 20.11 & 20.570 & 0.078  & 19.880 & 0.035  & 19.531 & 0.043  & 19.318 & 0.046 \\
GC185 & 00:53:41.32 & -42:58:27.280 & 20.50 & 21.155 & 0.185  & 20.372 & 0.048  & 19.933 & 0.054  & 19.746 & 0.066 \\
GC186 & 00:53:41.33 & -43:07:43.690 & 20.11 & 20.844 & 0.151  & 20.077 & 0.043  & 19.629 & 0.043  & 19.363 & 0.051 \\
GC187 & 00:53:41.36 & -42:54:08.297 & 20.47 & 20.891 & 0.077  & 20.243 & 0.027  & 19.950 & 0.039  & 19.792 & 0.049 \\
GC188 & 00:53:41.38 & -43:06:35.725 & 20.44 & 21.021 & 0.159  & 20.133 & 0.036  & 19.736 & 0.042  & 19.539 & 0.048 \\
GC194 & 00:53:41.56 & -42:53:26.012 & 20.00 & 20.537 & 0.061  & 19.779 & 0.027  & 19.421 & 0.038  & 19.199 & 0.039 \\
GC195 & 00:53:41.63 & -43:07:58.987 & 19.47 & 19.962 & 0.043  & 19.188 & 0.022  & 18.853 & 0.025  & 18.675 & 0.027 \\
GC196 & 00:53:41.69 & -42:58:21.465 & 20.26 & 20.859 & 0.096  & 20.099 & 0.036  & 19.650 & 0.042  & 19.435 & 0.044 \\
GC196 & 00:53:41.70 & -42:58:17.850 & 20.26 & 20.762 & 0.109  & 20.042 & 0.043  & 19.606 & 0.048  & 19.415 & 0.053 \\
GC197 & 00:53:41.70 & -42:56:31.335 & 20.06 & 20.465 & 0.072  & 19.849 & 0.026  & 19.481 & 0.037  & 19.255 & 0.039 \\
GC199 & 00:53:41.72 & -43:05:16.615 & 19.88 & 20.345 & 0.055  & 19.684 & 0.023  & 19.376 & 0.027  & 19.208 & 0.033 \\
GC200 & 00:53:41.73 & -43:03:25.670 & 19.10 & 19.312 & 0.033  & 18.833 & 0.023  & 18.670 & 0.026  & 18.582 & 0.028 \\
GC203 & 00:53:41.79 & -42:52:39.748 & 19.81 & 20.225 & 0.050  & 19.574 & 0.026  & 19.254 & 0.035  & 19.070 & 0.040 \\
GC204 & 00:53:41.79 & -43:09:40.585 & 19.53 & 19.965 & 0.044  & 19.291 & 0.023  & 99.000 & 0.027  & 18.889 & 0.028 \\
GC205 & 00:53:41.87 & -43:04:02.140 & 19.18 & 19.758 & 0.043  & 18.915 & 0.023  & 18.450 & 0.026  & 18.141 & 0.027 \\
GC207 & 00:53:41.92 & -43:04:21.652 & 99.00 & 18.748 & 0.030  & 18.067 & 0.021  & 17.822 & 0.025  & 17.710 & 0.024 \\
GC208 & 00:53:41.95 & -42:57:47.290 & 18.57 & 19.007 & 0.041  & 18.298 & 0.032  & 17.974 & 0.037  & 18.014 & 0.024 \\
GC208 & 00:53:41.96 & -42:57:42.680 & 18.57 & 19.003 & 0.051  & 18.314 & 0.036  & 99.000 & 0.040  & 17.749 & 0.035 \\
GC209 & 00:53:41.96 & -42:53:25.420 & 20.46 & 20.885 & 0.078  & 20.203 & 0.027  & 19.916 & 0.039  & 19.837 & 0.048 \\
GC210 & 00:53:41.96 & -42:58:09.840 & 19.49 & 20.059 & 0.057  & 19.148 & 0.033  & 18.667 & 0.038  & 18.327 & 0.035 \\
GC210 & 00:53:41.97 & -42:58:05.970 & 19.49 & 19.874 & 0.069  & 19.164 & 0.036  & 18.700 & 0.040  & 18.289 & 0.036 \\
GC211 & 00:53:41.97 & -42:54:44.325 & 19.26 & 19.818 & 0.041  & 18.940 & 0.026  & 18.492 & 0.035  & 18.209 & 0.033 \\
GC214 & 00:53:41.99 & -43:05:09.253 & 99.00 & 21.031 & 0.102  & 20.287 & 0.025  & 19.890 & 0.030  & 19.758 & 0.038 \\
GC215 & 00:53:42.01 & -42:56:47.095 & 18.36 & 99.000 & 99.00  & 18.063 & 0.039  & 17.681 & 0.090  & 17.464 & 0.083 \\
GC219 & 00:53:42.04 & -43:03:47.225 & 99.00 & 20.971 & 0.110  & 20.242 & 0.037  & 20.200 & 0.060  & 19.905 & 0.068 \\
GC224 & 00:53:42.11 & -42:58:04.330 & 19.97 & 20.304 & 0.088  & 19.699 & 0.039  & 19.208 & 0.044  & 18.718 & 0.040 \\
GC226 & 00:53:42.11 & -42:55:15.380 & 20.08 & 99.000 & 99.00  & 19.941 & 0.042  & 19.739 & 0.093  & 19.614 & 0.091 \\
GC228 & 00:53:42.12 & -43:05:45.805 & 99.00 & 20.511 & 0.162  & 20.345 & 0.066  & 20.102 & 0.056  & 99.000 & 99.00 \\
GC229 & 00:53:42.16 & -43:07:17.137 & 20.15 & 20.634 & 0.077  & 19.948 & 0.024  & 19.669 & 0.028  & 19.497 & 0.033 \\
GC230 & 00:53:42.17 & -42:58:46.630 & 18.86 & 19.372 & 0.053  & 18.504 & 0.036  & 18.094 & 0.040  & 17.779 & 0.035 \\
GC231 & 00:53:42.19 & -42:56:24.365 & 18.64 & 99.000 & 99.00  & 18.425 & 0.040  & 18.169 & 0.091  & 18.024 & 0.083 \\
GC232 & 00:53:42.19 & -43:07:02.545 & 18.77 & 19.299 & 0.044  & 18.501 & 0.030  & 18.129 & 0.033  & 17.955 & 0.037 \\
GC233 & 00:53:42.19 & -43:04:29.102 & 19.37 & 19.943 & 0.043  & 19.036 & 0.022  & 18.590 & 0.025  & 18.293 & 0.025 \\
GC234 & 00:53:42.22 & -42:59:00.200 & 20.35 & 99.000 & 99.00  & 20.053 & 0.040  & 19.591 & 0.047  & 19.222 & 0.050 \\
GC235 & 00:53:42.25 & -43:02:49.360 & 99.00 & 20.928 & 0.105  & 19.965 & 0.037  & 19.359 & 0.047  & 19.059 & 0.046 \\
GC236 & 00:53:42.26 & -43:03:51.313 & 99.00 & 21.133 & 0.110  & 20.219 & 0.024  & 19.730 & 0.030  & 19.390 & 0.035 \\
GC238 & 00:53:42.28 & -42:58:57.560 & 19.10 & 19.500 & 0.061  & 18.787 & 0.036  & 18.392 & 0.040  & 18.135 & 0.035 \\
GC239 & 00:53:42.28 & -42:56:59.205 & 19.03 & 99.000 & 99.00  & 18.976 & 0.050  & 18.804 & 0.098  & 18.626 & 0.091 \\
GC245 & 00:53:42.34 & -42:56:45.320 & 18.68 & 99.000 & 99.00  & 18.499 & 0.041  & 18.258 & 0.091  & 18.114 & 0.084 \\
GC246 & 00:53:42.34 & -42:53:00.645 & 20.25 & 20.826 & 0.116  & 99.000 & 0.040  & 19.493 & 0.091  & 19.182 & 0.085 \\
GC248 & 00:53:42.35 & -43:05:29.037 & 99.00 & 18.158 & 0.081  & 18.167 & 0.030  & 17.809 & 0.030  & 17.632 & 0.032 \\
GC250 & 00:53:42.37 & -43:08:37.300 & 20.80 & 21.366 & 0.162  & 20.612 & 0.046  & 20.240 & 0.054  & 20.038 & 0.075 \\
GC251 & 00:53:42.39 & -43:07:27.980 & 20.92 & 20.633 & 0.104  & 20.623 & 0.026  & 20.142 & 0.032  & 19.876 & 0.041 \\
GC252 & 00:53:42.40 & -42:53:39.920 & 20.17 & 20.655 & 0.107  & 19.913 & 0.041  & 19.551 & 0.092  & 19.358 & 0.087 \\
GC253 & 00:53:42.43 & -43:08:04.000 & 99.00 & 99.000 & 99.00  & 21.526 & 0.074  & 21.277 & 0.120  & 99.000 & 99.00 \\
GC255 & 00:53:42.50 & -43:05:44.950 & 19.64 & 20.028 & 0.064  & 19.307 & 0.022  & 18.863 & 0.025  & 18.559 & 0.026 \\
GC258 & 00:53:42.56 & -43:05:02.597 & 99.00 & 18.832 & 0.040  & 17.899 & 0.021  & 17.575 & 0.025  & 17.408 & 0.024 \\
GC259 & 00:53:42.56 & -43:03:28.615 & 99.00 & 19.837 & 0.053  & 19.219 & 0.028  & 18.924 & 0.031  & 18.754 & 0.031 \\
GC261 & 00:53:42.58 & -43:05:34.612 & 99.00 & 20.634 & 0.113  & 20.284 & 0.030  & 19.936 & 0.035  & 19.811 & 0.040 \\
GC262 & 00:53:42.61 & -43:04:33.762 & 99.00 & 21.029 & 0.088  & 20.205 & 0.023  & 19.742 & 0.029  & 19.488 & 0.045 \\
GC264 & 00:53:42.64 & -43:04:01.340 & 19.96 & 20.542 & 0.062  & 19.738 & 0.025  & 19.301 & 0.028  & 19.045 & 0.032 \\
GC265 & 00:53:42.65 & -42:55:59.015 & 17.63 & 99.000 & 99.00  & 17.394 & 0.041  & 17.100 & 0.091  & 16.949 & 0.084 \\
GC266 & 00:53:42.66 & -43:05:01.785 & 17.53 & 18.024 & 0.038  & 17.277 & 0.023  & 16.941 & 0.026  & 16.756 & 0.027 \\
GC267 & 00:53:42.66 & -43:03:09.430 & 99.00 & 20.280 & 0.120  & 20.437 & 0.075  & 20.101 & 0.075  & 19.569 & 0.077 \\
GC268 & 00:53:42.70 & -42:56:00.970 & 20.34 & 99.000 & 99.00  & 20.320 & 0.041  & 19.955 & 0.094  & 19.678 & 0.090 \\
GC271 & 00:53:42.72 & -43:05:00.220 & 99.00 & 99.000 & 99.00  & 21.030 & 0.048  & 20.926 & 0.067  & 20.723 & 0.130 \\
GC272 & 00:53:42.73 & -43:08:16.042 & 19.84 & 20.256 & 0.062  & 19.619 & 0.023  & 19.381 & 0.027  & 19.226 & 0.031 \\
GC273 & 00:53:42.77 & -42:58:04.870 & 19.99 & 20.573 & 0.089  & 19.694 & 0.038  & 19.323 & 0.043  & 19.109 & 0.042 \\
GC274 & 00:53:42.78 & -43:03:45.695 & 99.00 & 21.147 & 0.145  & 20.681 & 0.030  & 20.287 & 0.037  & 20.005 & 0.046 \\
GC275 & 00:53:42.80 & -43:10:41.750 & 19.26 & 19.791 & 0.055  & 19.107 & 0.033  & 18.874 & 0.035  & 18.609 & 0.043 \\
GC276 & 00:53:42.81 & -43:03:19.275 & 99.00 & 20.527 & 0.063  & 19.837 & 0.024  & 19.487 & 0.028  & 19.310 & 0.032 \\
GC277 & 00:53:42.82 & -42:59:13.970 & 18.39 & 18.829 & 0.050  & 18.108 & 0.036  & 17.790 & 0.040  & 17.598 & 0.035 \\
GC278 & 00:53:42.84 & -42:58:59.370 & 18.81 & 19.283 & 0.054  & 18.530 & 0.036  & 18.151 & 0.040  & 17.917 & 0.036 \\
GC279 & 00:53:42.83 & -43:03:41.380 & 20.39 & 20.978 & 0.084  & 20.065 & 0.024  & 19.701 & 0.029  & 19.486 & 0.036 \\
GC281 & 00:53:42.89 & -42:58:33.750 & 19.42 & 19.954 & 0.066  & 19.127 & 0.036  & 18.686 & 0.040  & 18.394 & 0.036 \\
GC283 & 00:53:42.90 & -43:04:56.412 & 99.00 & 19.336 & 0.046  & 18.495 & 0.023  & 18.084 & 0.026  & 17.852 & 0.027 \\
GC284 & 00:53:42.92 & -43:07:54.975 & 19.83 & 20.251 & 0.063  & 19.651 & 0.023  & 19.406 & 0.027  & 19.298 & 0.031 \\
GC286 & 00:53:42.96 & -42:58:56.140 & 18.58 & 19.061 & 0.052  & 18.304 & 0.036  & 17.949 & 0.040  & 17.705 & 0.035 \\
GC286 & 00:53:42.94 & -42:59:00.410 & 18.58 & 19.039 & 0.068  & 18.281 & 0.047  & 17.962 & 0.061  & 17.579 & 0.174 \\
GC289 & 00:53:43.07 & -42:57:15.630 & 18.37 & 18.782 & 0.050  & 18.175 & 0.037  & 17.910 & 0.041  & 17.730 & 0.036 \\
GC289 & 00:53:43.06 & -42:57:20.505 & 18.37 & 99.000 & 99.00  & 18.194 & 0.041  & 17.983 & 0.091  & 17.871 & 0.084 \\
GC290 & 00:53:43.06 & -43:06:45.435 & 99.00 & 21.275 & 0.128  & 20.593 & 0.027  & 20.222 & 0.033  & 20.072 & 0.046 \\
GC291 & 00:53:43.07 & -42:56:53.175 & 19.63 & 99.000 & 99.00  & 19.415 & 0.040  & 19.178 & 0.091  & 19.037 & 0.085 \\
GC295 & 00:53:43.11 & -42:57:03.115 & 19.87 & 99.000 & 99.00  & 19.565 & 0.039  & 19.145 & 0.090  & 18.933 & 0.084 \\
GC296 & 00:53:43.11 & -42:53:48.230 & 20.05 & 99.000 & 99.00  & 19.804 & 0.041  & 19.459 & 0.091  & 19.222 & 0.087 \\
GC297 & 00:53:43.13 & -43:08:06.673 & 19.23 & 19.741 & 0.049  & 18.940 & 0.022  & 18.581 & 0.025  & 18.399 & 0.026 \\
GC299 & 00:53:43.14 & -43:06:08.877 & 19.92 & 20.429 & 0.069  & 19.670 & 0.023  & 19.308 & 0.027  & 19.168 & 0.030 \\
GC300 & 00:53:43.19 & -43:00:42.485 & 19.71 & 20.183 & 0.060  & 19.525 & 0.029  & 19.192 & 0.035  & 19.047 & 0.045 \\
GC305 & 00:53:43.29 & -43:02:20.205 & 19.55 & 20.023 & 0.046  & 19.306 & 0.023  & 18.964 & 0.025  & 18.756 & 0.027 \\
GC307 & 00:53:43.31 & -43:05:04.680 & 20.12 & 20.740 & 0.085  & 19.830 & 0.023  & 19.368 & 0.027  & 19.077 & 0.030 \\
GC311 & 00:53:43.35 & -43:06:08.560 & 20.17 & 20.729 & 0.083  & 19.926 & 0.023  & 19.520 & 0.027  & 19.250 & 0.031 \\
GC312 & 00:53:43.36 & -43:04:08.225 & 19.97 & 20.470 & 0.073  & 19.713 & 0.023  & 19.336 & 0.027  & 19.151 & 0.031 \\
GC313 & 00:53:43.36 & -43:00:30.650 & 20.42 & 20.892 & 0.113  & 20.321 & 0.044  & 20.070 & 0.053  & 20.052 & 0.072 \\
GC314 & 00:53:43.37 & -42:57:58.020 & 19.42 & 19.926 & 0.062  & 19.209 & 0.036  & 18.969 & 0.041  & 18.840 & 0.040 \\
GC315 & 00:53:43.40 & -42:55:36.140 & 99.00 & 99.000 & 99.00  & 20.562 & 0.045  & 20.281 & 0.095  & 20.150 & 0.134 \\
GC316 & 00:53:43.43 & -42:59:26.500 & 19.92 & 20.393 & 0.081  & 19.692 & 0.040  & 19.367 & 0.046  & 19.168 & 0.048 \\
GC317 & 00:53:43.44 & -42:59:44.240 & 19.82 & 20.286 & 0.075  & 19.617 & 0.038  & 19.307 & 0.044  & 19.105 & 0.045 \\
GC319 & 00:53:43.49 & -42:58:26.260 & 19.62 & 20.107 & 0.069  & 19.364 & 0.036  & 19.073 & 0.043  & 18.834 & 0.041 \\
GC320 & 00:53:43.52 & -43:05:46.455 & 17.87 & 18.364 & 0.039  & 17.614 & 0.023  & 17.276 & 0.026  & 17.086 & 0.026 \\
GC321 & 00:53:43.53 & -42:58:37.970 & 19.97 & 20.477 & 0.084  & 19.670 & 0.038  & 19.284 & 0.043  & 19.021 & 0.044 \\
GC322 & 00:53:43.53 & -43:01:59.790 & 19.00 & 19.613 & 0.048  & 18.630 & 0.022  & 18.154 & 0.025  & 17.881 & 0.024 \\
GC325 & 00:53:43.57 & -43:03:56.548 & 19.48 & 20.030 & 0.055  & 19.228 & 0.023  & 18.817 & 0.026  & 18.631 & 0.027 \\
GC326 & 00:53:43.58 & -42:59:04.270 & 18.15 & 18.646 & 0.047  & 17.877 & 0.036  & 17.537 & 0.039  & 17.342 & 0.034 \\
GC328 & 00:53:43.62 & -42:56:20.570 & 20.38 & 99.000 & 99.00  & 20.122 & 0.043  & 19.899 & 0.094  & 19.727 & 0.091 \\
GC330 & 00:53:43.65 & -42:59:22.330 & 17.22 & 17.797 & 0.045  & 16.885 & 0.035  & 16.431 & 0.039  & 16.135 & 0.034 \\
GC331 & 00:53:43.68 & -42:59:10.670 & 18.88 & 19.272 & 0.063  & 18.536 & 0.043  & 18.095 & 0.044  & 17.845 & 0.040 \\
GC333 & 00:53:43.75 & -43:01:32.508 & 20.24 & 20.599 & 0.074  & 20.028 & 0.022  & 19.857 & 0.028  & 19.726 & 0.038 \\
GC339 & 00:53:43.90 & -43:08:06.315 & 19.30 & 19.855 & 0.051  & 19.002 & 0.023  & 18.583 & 0.026  & 18.295 & 0.025 \\
GC340 & 00:53:43.91 & -43:07:11.070 & 19.21 & 19.676 & 0.049  & 99.000 & 0.023  & 18.692 & 0.027  & 18.549 & 0.028 \\
GC341 & 00:53:43.93 & -42:53:18.925 & 19.58 & 99.000 & 99.00  & 19.316 & 0.039  & 19.018 & 0.091  & 18.775 & 0.116 \\
GC342 & 00:53:43.97 & -42:55:30.730 & 19.81 & 99.000 & 99.00  & 19.546 & 0.040  & 19.349 & 0.091  & 19.205 & 0.085 \\
GC347 & 00:53:44.05 & -43:09:40.055 & 20.09 & 20.615 & 0.090  & 19.785 & 0.025  & 19.413 & 0.030  & 19.189 & 0.030 \\
GC350 & 00:53:44.07 & -43:06:55.325 & 20.05 & 99.000 & 99.00  & 19.840 & 0.035  & 19.559 & 0.042  & 19.353 & 0.045 \\
GC354 & 00:53:44.15 & -43:08:55.683 & 19.84 & 20.353 & 0.067  & 19.561 & 0.023  & 19.187 & 0.027  & 18.952 & 0.029 \\
GC355 & 00:53:44.17 & -43:00:09.260 & 20.28 & 20.698 & 0.099  & 20.076 & 0.038  & 19.823 & 0.049  & 19.709 & 0.055 \\
GC357 & 00:53:44.19 & -42:56:57.320 & 18.49 & 99.000 & 99.00  & 18.176 & 0.041  & 17.794 & 0.091  & 17.486 & 0.084 \\
GC358 & 00:53:44.19 & -43:05:42.973 & 20.36 & 20.936 & 0.096  & 20.101 & 0.024  & 19.675 & 0.029  & 19.369 & 0.032 \\
GC361 & 00:53:44.28 & -42:55:44.770 & 19.10 & 99.000 & 99.00  & 18.848 & 0.039  & 18.605 & 0.090  & 18.452 & 0.083 \\
GC362 & 00:53:44.31 & -43:09:10.233 & 20.12 & 20.523 & 0.070  & 19.898 & 0.023  & 19.658 & 0.028  & 19.551 & 0.035 \\
GC365 & 00:53:44.36 & -42:56:32.630 & 17.17 & 99.000 & 99.00  & 16.891 & 0.042  & 16.642 & 0.092  & 16.477 & 0.084 \\
GC366 & 00:53:44.42 & -42:56:44.400 & 21.28 & 99.000 & 99.00  & 20.866 & 0.045  & 20.422 & 0.095  & 20.192 & 0.132 \\
GC367 & 00:53:44.43 & -43:00:37.578 & 19.43 & 19.850 & 0.051  & 19.201 & 0.022  & 18.932 & 0.026  & 18.783 & 0.028 \\
GC368 & 00:53:44.44 & -43:06:14.542 & 19.95 & 20.490 & 0.069  & 19.671 & 0.024  & 19.284 & 0.027  & 19.061 & 0.031 \\
GC371 & 00:53:44.46 & -43:07:52.703 & 20.40 & 20.925 & 0.095  & 20.131 & 0.024  & 19.780 & 0.029  & 19.589 & 0.034 \\
GC373 & 00:53:44.57 & -42:59:15.520 & 20.09 & 20.578 & 0.093  & 19.851 & 0.038  & 19.526 & 0.045  & 19.252 & 0.045 \\
GC374 & 00:53:44.59 & -43:01:20.995 & 18.99 & 19.444 & 0.044  & 18.776 & 0.023  & 18.490 & 0.026  & 18.330 & 0.027 \\
GC375 & 00:53:44.64 & -43:07:06.085 & 20.08 & 20.670 & 0.082  & 19.849 & 0.025  & 19.438 & 0.030  & 19.191 & 0.033 \\
GC377 & 00:53:44.66 & -42:56:36.385 & 99.00 & 99.000 & 99.00  & 21.282 & 0.051  & 20.963 & 0.109  & 20.644 & 0.150 \\
GC378 & 00:53:44.71 & -42:53:42.970 & 18.43 & 99.000 & 99.00  & 18.112 & 0.039  & 17.800 & 0.090  & 17.563 & 0.083 \\
GC379 & 00:53:44.86 & -43:09:09.253 & 18.66 & 19.182 & 0.042  & 18.359 & 0.022  & 17.955 & 0.025  & 17.718 & 0.026 \\
GC380 & 00:53:44.95 & -43:08:30.492 & 19.17 & 19.600 & 0.046  & 18.911 & 0.022  & 18.636 & 0.026  & 18.469 & 0.026 \\
GC383 & 00:53:45.15 & -43:06:39.352 & 20.09 & 20.523 & 0.072  & 19.892 & 0.024  & 19.587 & 0.029  & 19.429 & 0.035 \\
GC384 & 00:53:45.31 & -43:03:18.465 & 18.74 & 19.218 & 0.044  & 18.485 & 0.023  & 18.122 & 0.026  & 17.898 & 0.026 \\
GC385 & 00:53:45.35 & -43:10:35.350 & 18.95 & 19.297 & 0.043  & 18.694 & 0.030  & 18.449 & 0.035  & 18.278 & 0.038 \\
GC386 & 00:53:45.37 & -43:03:18.398 & 19.33 & 19.691 & 0.067  & 19.045 & 0.022  & 18.635 & 0.025  & 18.385 & 0.026 \\
GC391 & 00:53:45.47 & -42:53:46.175 & 19.68 & 99.000 & 99.00  & 19.435 & 0.039  & 19.231 & 0.092  & 19.133 & 0.086 \\
GC392 & 00:53:45.47 & -42:54:27.010 & 19.87 & 99.000 & 99.00  & 19.621 & 0.040  & 19.445 & 0.092  & 19.270 & 0.087 \\
GC393 & 00:53:45.47 & -43:09:10.808 & 19.82 & 20.245 & 0.061  & 19.514 & 0.023  & 19.224 & 0.026  & 19.054 & 0.029 \\
GC395 & 00:53:45.57 & -43:03:43.505 & 19.38 & 19.841 & 0.052  & 19.137 & 0.030  & 18.821 & 0.035  & 18.612 & 0.039 \\
GC406 & 00:53:46.80 & -43:07:44.935 & 17.43 & 99.000 & 99.00  & 17.101 & 0.027  & 16.872 & 0.033  & 16.718 & 0.033 \\
GC408 & 00:53:47.29 & -43:04:57.540 & 20.80 & 99.000 & 99.00  & 20.393 & 0.042  & 19.810 & 0.048  & 19.386 & 0.051 \\
GC417 & 00:53:38.83 & -42:51:48.878 & 99.00 & 20.000 & 0.051  & 19.281 & 0.025  & 18.925 & 0.035  & 18.746 & 0.034 \\
GC418 & 00:53:38.98 & -43:06:33.230 & 99.00 & 21.122 & 0.123  & 20.256 & 0.050  & 19.759 & 0.053  & 19.477 & 0.059 \\
GC420 & 00:53:39.73 & -43:02:15.805 & 99.00 & 20.851 & 0.096  & 20.157 & 0.038  & 19.800 & 0.043  & 19.683 & 0.048 \\
GC421 & 00:53:39.97 & -43:06:40.780 & 99.00 & 99.000 & 99.00  & 20.893 & 0.052  & 20.514 & 0.060  & 20.110 & 0.068 \\
GC422 & 00:53:40.22 & -43:08:14.245 & 99.00 & 20.678 & 0.086  & 19.986 & 0.037  & 19.707 & 0.042  & 19.535 & 0.043 \\
GC424 & 00:53:40.46 & -43:02:40.125 & 99.00 & 20.165 & 0.061  & 19.463 & 0.035  & 19.133 & 0.039  & 18.958 & 0.040 \\
GC425 & 00:53:40.64 & -43:04:37.820 & 99.00 & 21.053 & 0.115  & 20.327 & 0.039  & 20.018 & 0.048  & 19.913 & 0.060 \\
GC426 & 00:53:40.95 & -43:04:46.530 & 99.00 & 20.845 & 0.094  & 20.195 & 0.040  & 19.907 & 0.046  & 19.709 & 0.055 \\
GC429 & 00:53:41.53 & -43:08:10.253 & 99.00 & 19.501 & 0.038  & 18.371 & 0.022  & 17.949 & 0.025  & 17.724 & 0.024 \\
GC430 & 00:53:41.63 & -42:57:17.115 & 99.00 & 99.000 & 0.057  & 19.145 & 0.025  & 18.654 & 0.035  & 18.368 & 0.032 \\
GC430 & 00:53:41.64 & -42:57:12.100 & 99.00 & 20.019 & 0.067  & 19.106 & 0.036  & 18.684 & 0.040  & 18.310 & 0.036 \\
GC432 & 00:53:42.27 & -43:10:45.520 & 99.00 & 99.000 & 99.00  & 20.322 & 0.041  & 20.206 & 0.060  & 19.873 & 0.070 \\
GC434 & 00:53:42.52 & -42:56:28.050 & 99.00 & 99.000 & 99.00  & 18.573 & 0.039  & 18.181 & 0.090  & 17.909 & 0.083 \\
GC435 & 00:53:42.57 & -42:57:20.320 & 99.00 & 99.000 & 99.00  & 19.675 & 0.040  & 19.302 & 0.091  & 19.011 & 0.085 \\
GC436 & 00:53:42.79 & -42:58:27.990 & 99.00 & 19.514 & 0.055  & 18.712 & 0.036  & 18.291 & 0.039  & 18.013 & 0.035 \\
GC437 & 00:53:43.02 & -42:54:23.570 & 99.00 & 99.000 & 99.00  & 20.334 & 0.043  & 19.925 & 0.094  & 19.693 & 0.089 \\
GC439 & 00:53:43.13 & -42:55:29.620 & 99.00 & 99.000 & 99.00  & 19.160 & 0.040  & 18.816 & 0.091  & 18.585 & 0.084 \\
GC440 & 00:53:43.36 & -42:51:37.810 & 99.00 & 19.905 & 0.060  & 19.106 & 0.036  & 18.859 & 0.041  & 18.609 & 0.040 \\
GC442 & 00:53:44.62 & -42:53:17.920 & 99.00 & 99.000 & 99.00  & 19.795 & 0.042  & 19.573 & 0.091  & 19.312 & 0.125 \\
GC444 & 00:53:45.38 & -42:53:46.655 & 99.00 & 99.000 & 99.00  & 20.253 & 0.040  & 20.087 & 0.094  & 20.076 & 0.132 \\
GC445 & 00:53:38.70 & -43:01:56.695 & 99.00 & 21.169 & 0.120  & 20.335 & 0.037  & 99.000 & 0.043  & 19.715 & 0.051 \\
GC446 & 00:53:39.36 & -43:04:51.005 & 99.00 & 20.397 & 0.070  & 19.409 & 0.033  & 18.921 & 0.038  & 18.589 & 0.036 \\
GC448 & 00:53:39.58 & -42:57:59.055 & 99.00 & 21.371 & 0.141  & 20.497 & 0.039  & 20.030 & 0.047  & 19.800 & 0.062 \\
GC449 & 00:53:39.60 & -43:00:43.600 & 99.00 & 20.836 & 0.095  & 99.000 & 0.036  & 19.524 & 0.042  & 19.146 & 0.042 \\
GC451 & 00:53:39.70 & -42:59:48.505 & 99.00 & 20.427 & 0.071  & 19.718 & 0.035  & 19.364 & 0.040  & 19.157 & 0.043 \\
GC452 & 00:53:39.85 & -42:58:51.640 & 99.00 & 21.691 & 0.190  & 20.931 & 0.040  & 20.502 & 0.062  & 20.435 & 0.096 \\
GC453 & 00:53:39.91 & -42:58:16.590 & 99.00 & 19.552 & 0.048  & 18.681 & 0.034  & 18.238 & 0.038  & 17.929 & 0.036 \\
GC457 & 00:53:40.46 & -42:57:32.200 & 99.00 & 20.657 & 0.083  & 19.745 & 0.033  & 19.307 & 0.039  & 19.000 & 0.038 \\
GC458 & 00:53:40.51 & -42:55:49.650 & 99.00 & 21.495 & 0.125  & 21.130 & 0.032  & 21.021 & 0.058  & 99.000 & 99.00 \\
GC459 & 00:53:41.18 & -42:56:13.925 & 99.00 & 99.000 & 99.00  & 21.426 & 0.037  & 20.967 & 0.056  & 20.673 & 0.094 \\
GC460 & 00:53:41.36 & -43:06:08.840 & 99.00 & 21.021 & 0.160  & 20.152 & 0.039  & 19.684 & 0.043  & 19.440 & 0.048 \\
GC461 & 00:53:41.60 & -43:03:45.198 & 99.00 & 21.172 & 0.110  & 20.511 & 0.029  & 20.154 & 0.036  & 20.195 & 0.056 \\
GC462 & 00:53:41.73 & -43:03:25.670 & 99.00 & 19.312 & 0.033  & 18.833 & 0.023  & 18.670 & 0.026  & 18.582 & 0.028 \\
GC463 & 00:53:41.82 & -42:58:28.880 & 99.00 & 20.172 & 0.062  & 19.328 & 0.034  & 18.850 & 0.038  & 18.592 & 0.038 \\
GC463 & 00:53:41.83 & -42:58:25.380 & 99.00 & 20.161 & 0.074  & 19.342 & 0.038  & 18.890 & 0.042  & 18.601 & 0.037 \\
GC464 & 00:53:41.83 & -42:55:30.118 & 99.00 & 99.000 & 99.00  & 20.867 & 0.030  & 20.522 & 0.046  & 20.198 & 0.059 \\
GC465 & 00:53:42.01 & -43:06:54.745 & 99.00 & 99.000 & 99.00  & 21.258 & 0.040  & 20.889 & 0.051  & 20.397 & 0.166 \\
GC467 & 00:53:42.07 & -42:58:33.510 & 99.00 & 20.163 & 0.074  & 19.261 & 0.040  & 18.837 & 0.045  & 18.560 & 0.044 \\
GC468 & 00:53:42.08 & -42:55:47.625 & 99.00 & 99.000 & 99.00  & 21.269 & 0.051  & 20.968 & 0.107  & 20.575 & 0.149 \\
GC470 & 00:53:42.21 & -42:54:52.920 & 99.00 & 99.000 & 99.00  & 20.735 & 0.055  & 20.161 & 0.065  & 99.000 & 99.00 \\
GC470 & 00:53:42.21 & -42:54:56.965 & 99.00 & 99.000 & 99.00  & 20.883 & 0.047  & 20.677 & 0.097  & 99.000 & 99.00 \\
GC470 & 00:53:42.19 & -42:55:00.365 & 99.00 & 99.000 & 99.00  & 20.749 & 0.049  & 20.332 & 0.119  & 19.845 & 0.094 \\
GC471 & 00:53:42.81 & -42:57:24.355 & 99.00 & 99.000 & 99.00  & 19.787 & 0.046  & 19.651 & 0.096  & 19.511 & 0.094 \\
GC473 & 00:53:42.87 & -43:00:05.150 & 99.00 & 99.000 & 99.00  & 20.736 & 0.053  & 20.341 & 0.065  & 20.170 & 0.083 \\
GC474 & 00:53:42.99 & -42:58:20.550 & 99.00 & 20.768 & 0.108  & 20.031 & 0.047  & 19.792 & 0.061  & 19.647 & 0.061 \\
GC476 & 00:53:43.15 & -42:57:45.120 & 99.00 & 20.034 & 0.064  & 19.417 & 0.036  & 19.105 & 0.042  & 18.938 & 0.042 \\
GC478 & 00:53:43.29 & -42:59:59.790 & 99.00 & 19.710 & 0.059  & 19.007 & 0.037  & 18.697 & 0.040  & 18.569 & 0.040 \\
GC479 & 00:53:44.01 & -42:59:55.490 & 99.00 & 21.511 & 0.191  & 21.003 & 0.048  & 20.840 & 0.083  & 99.000 & 99.00 \\
GC490 & 00:53:39.16 & -43:09:59.860 & 99.00 & 21.343 & 0.138  & 20.487 & 0.037  & 20.068 & 0.045  & 19.784 & 0.057 \\
GC493 & 00:53:39.86 & -43:07:06.255 & 99.00 & 21.269 & 0.131  & 20.427 & 0.038  & 20.062 & 0.046  & 19.795 & 0.059 \\
GC496 & 00:53:39.99 & -42:57:33.220 & 99.00 & 20.602 & 0.079  & 19.927 & 0.034  & 19.609 & 0.042  & 19.436 & 0.046 \\
GC498 & 00:53:40.09 & -43:05:46.335 & 99.00 & 21.724 & 0.191  & 20.753 & 0.040  & 20.248 & 0.046  & 19.857 & 0.145 \\
GC499 & 00:53:40.12 & -43:03:48.025 & 99.00 & 21.678 & 0.183  & 20.789 & 0.041  & 20.326 & 0.055  & 20.046 & 0.064 \\
GC500 & 00:53:40.24 & -43:01:41.413 & 99.00 & 99.000 & 99.00  & 20.691 & 0.035  & 20.372 & 0.046  & 20.187 & 0.073 \\
GC502 & 00:53:40.45 & -43:08:58.230 & 99.00 & 99.000 & 99.00  & 21.348 & 0.080  & 20.867 & 0.089  & 99.000 & 99.00 \\
GC504 & 00:53:40.60 & -43:08:53.535 & 99.00 & 20.844 & 0.098  & 20.365 & 0.040  & 20.176 & 0.048  & 20.123 & 0.079 \\
GC505 & 00:53:40.85 & -43:01:55.775 & 99.00 & 20.811 & 0.096  & 20.011 & 0.038  & 19.632 & 0.043  & 19.367 & 0.042 \\
GC507 & 00:53:40.87 & -42:55:56.373 & 99.00 & 20.463 & 0.057  & 19.805 & 0.026  & 19.487 & 0.037  & 19.286 & 0.037 \\
GC510 & 00:53:41.18 & -42:52:56.040 & 99.00 & 99.000 & 99.00  & 20.842 & 0.046  & 20.583 & 0.097  & 20.350 & 0.140 \\
GC520 & 00:53:41.81 & -42:58:37.990 & 99.00 & 21.146 & 0.124  & 20.228 & 0.036  & 19.734 & 0.042  & 19.548 & 0.052 \\
GC520 & 00:53:41.82 & -42:58:34.650 & 99.00 & 21.304 & 0.169  & 20.376 & 0.046  & 19.803 & 0.049  & 19.611 & 0.057 \\
GC521 & 00:53:41.92 & -42:53:15.757 & 99.00 & 20.777 & 0.070  & 19.965 & 0.026  & 19.489 & 0.037  & 19.243 & 0.038 \\
GC523 & 00:53:42.21 & -43:09:21.647 & 99.00 & 21.265 & 0.124  & 20.628 & 0.026  & 20.395 & 0.038  & 20.239 & 0.050 \\
GC527 & 00:53:42.64 & -42:58:20.410 & 99.00 & 20.576 & 0.094  & 20.016 & 0.040  & 19.776 & 0.049  & 19.600 & 0.054 \\
GC528 & 00:53:43.22 & -42:55:06.610 & 99.00 & 99.000 & 99.00  & 20.024 & 0.040  & 19.665 & 0.091  & 19.387 & 0.087 \\
GC529 & 00:53:43.40 & -43:08:03.098 & 99.00 & 99.000 & 99.00  & 20.407 & 0.027  & 19.847 & 0.029  & 19.574 & 0.036 \\
GC529 & 00:53:43.37 & -43:08:02.985 & 99.00 & 21.050 & 0.106  & 20.493 & 0.034  & 20.193 & 0.046  & 20.325 & 0.081 \\
GC530 & 00:53:43.50 & -43:02:59.957 & 99.00 & 21.367 & 0.139  & 20.528 & 0.025  & 20.196 & 0.034  & 19.956 & 0.041 \\
GC531 & 00:53:43.65 & -43:01:40.450 & 99.00 & 99.000 & 99.00  & 20.528 & 0.039  & 20.019 & 0.045  & 19.609 & 0.051 \\
GC532 & 00:53:43.65 & -42:52:57.160 & 99.00 & 99.000 & 99.00  & 20.850 & 0.049  & 20.466 & 0.096  & 20.166 & 0.134 \\
GC533 & 00:53:43.67 & -43:10:04.720 & 99.00 & 20.952 & 0.097  & 20.253 & 0.023  & 19.924 & 0.029  & 19.725 & 0.039 \\
GC534 & 00:53:43.69 & -42:53:40.220 & 99.00 & 99.000 & 99.00  & 20.812 & 0.045  & 20.566 & 0.118  & 20.379 & 0.142 \\
GC537 & 00:53:43.75 & -43:02:53.020 & 99.00 & 99.000 & 99.00  & 20.666 & 0.035  & 20.159 & 0.047  & 19.874 & 0.185 \\
GC541 & 00:53:44.09 & -43:02:34.620 & 99.00 & 21.529 & 0.156  & 20.731 & 0.026  & 20.349 & 0.034  & 20.173 & 0.048 \\
GC543 & 00:53:44.57 & -43:04:44.977 & 99.00 & 21.148 & 0.111  & 20.483 & 0.025  & 20.169 & 0.032  & 20.078 & 0.046 \\
GC544 & 00:53:45.15 & -43:09:57.920 & 99.00 & 99.000 & 99.00  & 21.148 & 0.039  & 20.728 & 0.057  & 20.348 & 0.063 \\
GC551 & 00:53:46.03 & -43:03:14.890 & 99.00 & 21.020 & 0.115  & 20.639 & 0.038  & 20.311 & 0.054  & 19.942 & 0.188 \\
GC552 & 00:53:46.51 & -43:10:40.580 & 99.00 & 99.000 & 99.00  & 21.273 & 0.116  & 20.627 & 0.091  & 19.981 & 0.110 \\
GC557 & 00:53:47.39 & -43:07:20.545 & 99.00 & 21.369 & 0.134  & 20.659 & 0.035  & 20.361 & 0.064  & 20.020 & 0.060 \\
GC572 & 00:53:38.49 & -43:06:31.340 & 99.00 & 21.315 & 0.134  & 20.419 & 0.038  & 19.956 & 0.043  & 19.578 & 0.132 \\
GC575 & 00:53:39.19 & -43:08:43.525 & 99.00 & 20.270 & 0.066  & 19.760 & 0.033  & 19.560 & 0.040  & 19.519 & 0.047 \\
GC577 & 00:53:39.65 & -43:08:49.170 & 99.00 & 19.668 & 0.049  & 19.013 & 0.033  & 18.833 & 0.038  & 18.729 & 0.036 \\
GC578 & 00:53:39.69 & -42:58:15.050 & 99.00 & 21.443 & 0.148  & 20.582 & 0.036  & 20.127 & 0.048  & 19.845 & 0.061 \\
GC581 & 00:53:39.91 & -42:58:05.550 & 99.00 & 99.000 & 0.111  & 20.498 & 0.046  & 20.154 & 0.052  & 99.000 & 99.00 \\
GC582 & 00:53:40.08 & -43:04:01.475 & 99.00 & 21.226 & 0.125  & 20.272 & 0.037  & 19.813 & 0.043  & 19.480 & 0.048 \\
GC583 & 00:53:40.15 & -42:56:51.933 & 99.00 & 21.185 & 0.098  & 20.438 & 0.028  & 20.117 & 0.041  & 19.887 & 0.051 \\
GC584 & 00:53:40.60 & -43:10:01.895 & 99.00 & 20.381 & 0.096  & 19.554 & 0.034  & 19.180 & 0.038  & 18.949 & 0.038 \\
GC585 & 00:53:40.94 & -43:02:42.580 & 99.00 & 20.707 & 0.088  & 20.039 & 0.037  & 19.706 & 0.042  & 19.680 & 0.049 \\
GC586 & 00:53:41.10 & -43:06:11.895 & 99.00 & 21.235 & 0.128  & 20.360 & 0.037  & 19.893 & 0.042  & 19.593 & 0.051 \\
GC587 & 00:53:41.34 & -43:03:09.875 & 99.00 & 20.545 & 0.081  & 19.757 & 0.037  & 19.315 & 0.042  & 19.103 & 0.040 \\
GC588 & 00:53:42.01 & -42:54:00.555 & 99.00 & 20.759 & 0.116  & 20.083 & 0.040  & 19.778 & 0.092  & 19.750 & 0.093 \\
GC589 & 00:53:42.55 & -42:54:53.450 & 99.00 & 99.000 & 99.00  & 19.986 & 0.040  & 19.492 & 0.091  & 19.259 & 0.086 \\
GC590 & 00:53:43.13 & -42:52:33.985 & 99.00 & 20.339 & 0.088  & 19.632 & 0.040  & 19.342 & 0.092  & 19.109 & 0.086 \\
GC593 & 00:53:43.47 & -42:55:28.130 & 99.00 & 99.000 & 99.00  & 19.979 & 0.040  & 19.700 & 0.091  & 19.521 & 0.088 \\
GC594 & 00:53:43.67 & -43:00:41.210 & 99.00 & 20.756 & 0.071  & 20.071 & 0.023  & 19.783 & 0.029  & 19.647 & 0.037 \\
GC595 & 00:53:43.85 & -43:05:13.243 & 99.00 & 20.722 & 0.107  & 19.935 & 0.030  & 19.501 & 0.035  & 19.328 & 0.038 \\
GC597 & 00:53:44.44 & -43:07:08.203 & 99.00 & 21.067 & 0.110  & 20.220 & 0.026  & 19.805 & 0.030  & 19.579 & 0.037 \\
GC598 & 00:53:44.64 & -43:08:52.450 & 99.00 & 20.569 & 0.077  & 19.759 & 0.025  & 19.344 & 0.028  & 19.008 & 0.031 \\
GC599 & 00:53:44.70 & -43:01:05.470 & 99.00 & 20.810 & 0.087  & 19.941 & 0.023  & 19.499 & 0.027  & 19.289 & 0.032 \\
GC603 & 00:53:45.41 & -43:02:59.222 & 99.00 & 19.779 & 0.097  & 20.037 & 0.022  & 19.675 & 0.028  & 19.406 & 0.034 \\
GC605 & 00:53:47.60 & -43:05:06.025 & 99.00 & 99.000 & 99.00  & 19.412 & 0.036  & 19.067 & 0.042  & 18.787 & 0.045 \\
\end{longtable}

\end{document}